\documentclass[useAMS,usenatbib]{mnras}

 \usepackage{amssymb}
 \usepackage{graphicx}
 \usepackage{epstopdf}
 \usepackage[colorinlistoftodos]{todonotes}
 \usepackage{natbib}
 \usepackage{float}
 \usepackage{amsmath}
 \usepackage{dirtytalk}
 \usepackage{booktabs}
 \usepackage{color,soul}
 \usepackage{placeins}
\usepackage{hyperref}

\newcommand{\pd}[2]{\frac{\partial #1}{\partial #2}} 
 
\newcommand{\nhat}{\boldsymbol{\hat{n}}}

\title[Gravitational and L-ISW bispectrum of 21cm signal]{The gravitational and lensing-ISW bispectrum of 21cm radiation}

\author[C. J. Schmit et al.]{
Claude J. Schmit,\thanks{E-mail: c.schmit13@imperial.ac.uk}
Alan F. Heavens,
Jonathan R. Pritchard
\\
Imperial Centre for Inference and Cosmology, Imperial College London, Prince Consort Road, London SW7 2AZ, UK
}

\date{Accepted XXX. Received YYY; in original form ZZZ}

\pubyear{2018}

\begin{document}
\label{firstpage}
\pagerange{\pageref{firstpage}--\pageref{lastpage}}
\maketitle

\begin{abstract}
Cosmic Microwave Background experiments from COBE to \textit{Planck}, have launched cosmology into an era of precision science, where many cosmological parameters are now determined to the percent level. 
Next generation telescopes, focussing on the cosmological 21cm signal from neutral hydrogen, will probe enormous volumes in the low-redshift Universe, and have the potential to determine dark energy properties and test modifications of Einstein's gravity. 
We study the 21cm bispectrum due to gravitational collapse as well as the contribution by line of sight perturbations in the form of the lensing-ISW bispectrum at low-redshifts ($z \sim 0.35-3$), targeted by upcoming neutral hydrogen intensity mapping experiments.
We compute the expected bispectrum amplitudes and use a Fisher forecast model to compare power spectrum and bispectrum observations of intensity mapping surveys by CHIME, MeerKAT and SKA-mid.
We find that combined power spectrum and bispectrum observations have the potential to decrease errors on the cosmological parameters by an order of magnitude compared to \textit{Planck}.
Finally, we compute the contribution of the lensing-ISW bispectrum, and find that, unlike for the cosmic microwave background analyses, it can safely be ignored for 21cm bispectrum observations.
\end{abstract}



\section{Introduction}
 
Since its discovery, the Cosmic Microwave Background (CMB) has proven to be a rich seam of cosmological information and has propelled cosmology into an age of precision science.
Over the last three decades, experiments such as COBE \citep{Smoot1992}, WMAP \citep{Bennett2003,Bennett2013}, and Planck \citep{Planck2016a} have measured the CMB to an astonishing degree of accuracy, and its information is routinely combined with various other cosmological probes such as weak lensing, galaxy clustering and Type-1a supernovae.
This great effort has allowed us to constrain many of the parameters of the geometrically flat, cold dark matter model with a cosmological constant ($\Lambda$CDM) to the percent level.

Although the \textit{Planck} data favours a simple six parameter model over other models \citep{Heavens2017}, there remain tensions between the CMB measurements and local direct measurements of the Hubble parameter, $h$, \citep{Bennett2014,Riess2016,Riess2018a,Riess2018b}, as well as low-redshift weak lensing measurements, which find slightly less matter clumping than expected from extrapolating the CMB findings  \citep{Heymans2013, MacCrann2015, Raveri2016, Joudaki2017, Kohlinger2017}.
These tensions can arise if the assumed cosmological model is wrong since the CMB photons principally reveal the conditions of the Universe at the time of recombination at a relatively thin redshift slice at $z \simeq 1100$, when the Universe was matter dominated.
Additional probes along the line of sight are required to give the full 3-dimensional context for the evolution of the Universe and study the evolution of low-redshift phenomena, such as dark energy.
Galaxy surveys, such as the 2dF Galaxy Redshift survey \citep{Colless2001}, BOSS \citep{Anderson2012}, and SDSS \citep{Ahn2014}, are one such probe which determine the cosmological parameters by mapping the positions of galaxies in the sky and realizing that they are biased tracers of the underlying dark matter distribution.
These surveys thus relate the galaxy power spectrum directly to the matter power spectrum from which the parameters can be determined.
Weak lensing surveys, such as CFHTLenS \citep{Heymans2012}, KiDS \citep{DeJong2013} and DES \citep{Jarvis2016}, present another low-redshift observation that complements the CMB observations, as the reconstructed lensing potential is directly related to the gravitational potential of the Universe.
For both galaxy redshift surveys and weak lensing surveys it is crucial to obtain large galaxy samples by probing the largest observational volumes possible.
One of the main difficulties for these surveys is to determine the redshift information of galaxies in their sample, as the largest volumes are attained by rapid photometry of the sources.
Imprecise redshift information effectively blurs the radial information of the galaxies in the sample and propagates as a systematic error into the analysis.

Recently, much interest has been given to the potential that the 21cm spin-flip transition of the neutral hydrogen ground state has as a new low-redshift probe of the Universe \citep{Furlanetto2006, Pritchard2012}.
Due to the inherent relation between the observed frequency of the 21cm signal and the redshift at which it was emitted, the signal readily provides spectroscopic redshifts and hence much more precise 3D information about the Universe.
Much of the recent attention is due to the advent of next generation radio observatories (SKA\footnote{\url{https://www.skatelescope.org/}}, LOFAR \citep{Patil2017}, MWA \citep{Dillon2015}, HERA \citep{DeBoer2017}, CHIME\footnote{\url{https://chime-experiment.ca/}}, TIANLAI \citep{Chen2015}, BINGO \citep{Battye2016}), which predominantly target the redshifted 21cm signal throughout cosmic history, back as far as the epoch of reionization (EoR) and potentially the late stages of the dark ages. 
After the EoR, when most of the neutral hydrogen in the Universe has been ionized, the remaining atomic hydrogen resides mainly within self-shielded damped Lyman-$\alpha$ (DLA) systems inside galaxies and galaxy clusters. 
Intensity mapping (IM) experiments such as CHIME, BINGO, and TIANLAI integrate the 21cm emission of unresolved clouds of hydrogen gas within a given frequency bin.
This technique allows for large volume surveys with precise redshift information.
Foregrounds limit the sensitivity to the signal at all frequencies even after foregrounds have been statistically separated and removed, and therefore the statistical analysis of fluctuations in the 21cm brightness temperature is expected to hold the most potential for a detection of the signal.
IM thus provides CMB-like maps of the 21cm brightness temperature fluctuations in each frequency bin which can be similarly analysed for the power spectrum and bispectrum of the signal.
\cite{Bull2015} have thoroughly examined the information gained from power spectrum observation of an extensive list of 21cm IM experiments and find competitive percent level forecasts on the cosmological parameters. 
Theoretical predictions of the 21cm bispectrum due to primordial non-Gaussianities (PNG) and non-linear gravity collapse (NLG) \citep{Pillepich2007} give promising predictions of the signal to noise of high redshift bispectrum detections. 
However, the post-EoR 21cm signal is expected to be highly non-Gaussian as neutral hydrogen traces the galaxy population at late times. 
Due to this highly non-Gaussian field, the power spectrum cannot probe the full information content of the field and much of the low-redshift information should reside in these higher order statistics \citep{Repp2015}.
We thus evaluate the model for the 21cm bispectrum at low-redshifts and compute the Fisher forecasts combining power spectrum and bispectrum information.

We examine another physical effect which can lead to a non-zero bispectrum, the correlation between lensing and the integrated Sachs-Wolfe effect. 
As the 21cm emission travels towards our telescopes, it traverses the intergalactic medium (IGM) and is subjected to the gravitational effects of the intervening matter. 
Matter fluctuations act as gravitational lenses on the 21cm photons, whose paths get distorted by their presence. 
This effect should be noticeable through the statistical distribution of the 21cm photons on the sky.
In addition to this, as the Universe evolves into an acceleration-dominated era at low-redshifts ($z \lesssim 2$), the growth of structure lags behind the accelerated expansion of space. 
This effect causes the gravitational potentials of galaxy clusters to decrease in amplitude over time, resulting in a boost in energy for photons traversing those potentials.
This late-time integrated Sachs-Wolfe effect (ISW) once again distorts the intensity distribution of photons in a survey volume.
Cross-correlations between these two lines of sight effects improve cosmological parameter constraints from lensing surveys on the 10 percent level on large scales as shown by \cite{Zieser2016}.
Most importantly, however, ignoring the lensing-ISW (LISW) effect has been shown to bias CMB parameter inferences \citep{Kim2013}, and will at some level bias 21cm bispectrum observations. 
We compute both the LISW bispectrum and the bias resulting from neglecting it from upcoming IM experiments.

This paper is organised as follows; In section \ref{sec: 21cm Signal Model} we will introduce the 21cm signal model we use throughout, and write down the angular power spectrum.
In section \ref{sec: bispectrum}, we revisit the 21cm bispectrum from \cite{Pillepich2007} and include a low-redshift 21cm signal model. 
We also discuss the effects of lensing, the ISW effect, and the angular LISW bispectrum.
We then compute both the 21cm bispectrum and the LISW bispectrum for all triangle configurations at $z = 1$.
In section \ref{sec: instr}, we discuss upcoming intensity mapping experiments able to detect the 21cm bispectrum, and discuss foregrounds and noise.
Section \ref{sec: stats} introduces our forecast model and determines the expected signal to noise for a LISW bispectrum detection as well as the bias introduced when neglecting it.
Finally, we present and discuss the results of the parameter forecasts in section \ref{sec: discussion}, before we summarize our findings in section \ref{sec: conclusion}.
Throughout this paper, we assume a six parameter $\Lambda$CDM cosmology with fiducial values $(\Omega_\text{b} h^2, \Omega_\text{CDM}h^2, \Omega_\Lambda, h, 10^9 \times A_{\rm s}, n_{\rm s}) = (0.022,0.127,0.684,0.67,1.562,0.962)$.

\section{The 21cm signal}\label{sec: 21cm Signal Model}

In this section we will discuss the model for the 21cm brightness temperature evolution used in this paper.

\subsection{Brightness temperature fluctuations}
The 21cm signal originates from the hyperfine ground state transition in the hydrogen atom.
Its strength is governed by the relative abundance of HI atoms in the excited, triplet $(1)$, state relative to the non-excited, singlet $(0)$, state, parametrised through the spin temperature, $T_{\rm S}$,
\begin{equation}
\frac{n_1}{n_0} = \frac{g_1}{g_0}\exp\left(-\frac{T_*}{T_{\rm S}}\right),
\end{equation}
where $T_* = h\nu_{21}/k_{\rm B} \approx 68\text{mK}$, $g_i$ is the statistical weight of the energy level $i$, $g_1/g_0 = 3$, and $T_{\rm S} \gg T_*$.
The intensity of the signal on the sky is then measured, and we model the signal in terms of its brightness temperature, which relates to the signal intensity via the Rayleigh-Jeans formula, $T_{\rm b}(\nu)\approx I_\nu c^2/2k_{\rm B}\nu^2$. 
Generally, the 21cm signal is measured using the CMB as a background, 
\begin{equation}
T_{\rm b}(z) = \frac{T_{\rm S} - T_\gamma(z)}{1+z}\tau,
\end{equation}
where $T_\gamma(z)$ denotes the CMB temperature at redshift $z$ and $\tau$ is the optical depth through a cloud of neutral hydrogen.

\begin{figure}
\includegraphics[width=0.45\textwidth]{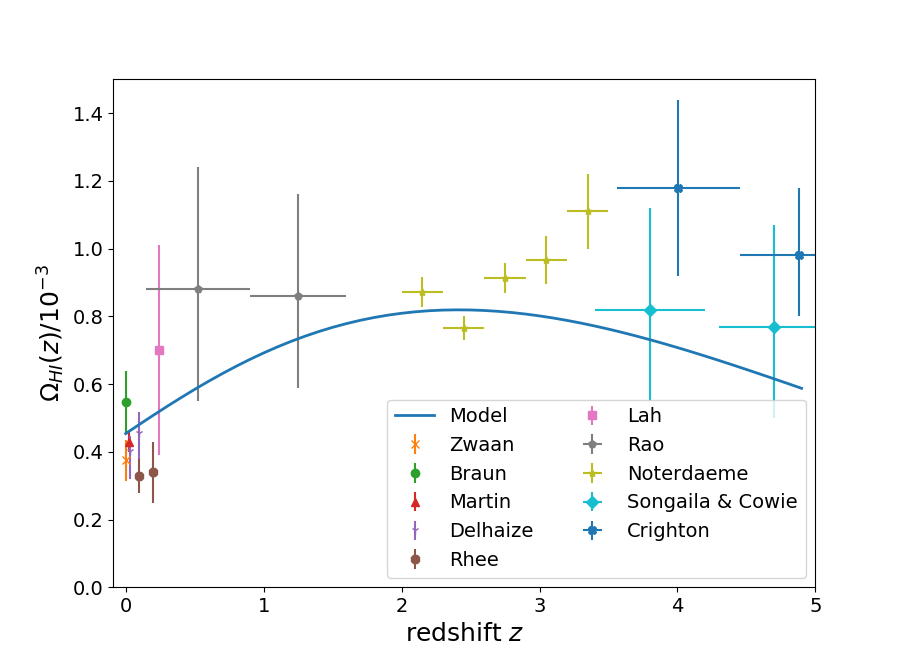}
\caption{Comparison between our analytic model for the HI density, $\tilde{\Omega}_\text{HI}$, as a function of redshift with current measurements. Included are the results from \protect\cite{Zwaan2015, Braun2012, Martin2010, Delhaize2013, Rhee2013, Lah2007, Rao2006, Noterdaeme2012, Songaila2010, Crighton2015}. See \protect\cite{Crighton2015} for full data list.}
\label{Figure: omegaHI}
\end{figure}

The spin temperature and thus $T_{\rm b}$ depend on the underlying HI density field as well as astrophysical effects, such that the brightness temperature can be split into a homogeneous and a fluctuating part,
\begin{equation}
T_{\rm b}(z) = \delta \bar{T}_{\rm b}(z) \left[1 + \delta_{\text{HI}}(z)\right].
\end{equation}
In the context of intensity mapping, we follow the model of \cite{Bull2015} and focus on the mean 21cm signal that is emitted by localised clumps of HI gas within galaxies and galaxy clusters for which the average brightness temperature over the sky can be approximated as \citep{Santos2015}
\begin{equation}\label{eq: temp model}
\delta \bar{T}_{\rm b} \approx 566h \left[\frac{H_0}{H(z)}\right]\left[\frac{\tilde{\Omega}_\text{HI}(z)}{0.003}\right](1+z)^2 \mu\text{K}.
\end{equation}
Here, $\tilde{\Omega}_\text{HI}$ is the density of HI atoms in units of the current critical density, 
\begin{equation}
\tilde{\Omega}_\text{HI}(z) \equiv \rho_\text{HI}(z)/\rho_{\rm c,0},
\end{equation}
with a critical density today, $\rho_{\rm c,0} = 3H^2_0/8\pi G$.
The density of neutral hydrogen is related to the mass of the dark matter halos in the Universe,
\begin{equation}
\rho_\text{HI}(z)=\int_{M_\text{min}}^{M_\text{max}}dM \frac{dn}{dm} M_\text{HI}(M),
\end{equation}
where $dn/dm$ is the halo mass function, for which we use a simple Sheth-Tormen implementation. 
Following \cite{Bagla2010}, we adopt a lower cutoff for the mass range containing HI gas, to correspond to a circular halo velocity of 30 km/s, meaning that halos with a lower circular velocity do not contain any HI gas.
Typically, neutral hydrogen can be expected in star forming halos, and gas in halos with circular velocities of larger than 60 km/s can be expected to form stars. 
Additionally, self-shielding damped Lyman-$\alpha$ systems can be found in lower mass halos, justifying a somewhat lower velocity cut-off.
The HI mass density is measured locally using 21cm emission \citep{Zwaan2005, Martin2010} and at higher redshifts via damped Lyman-$\alpha$ systems, as they trace the HI distribution after the EoR \citep{Prochaska2005}.
\cite{Crighton2015} summarize recent measurements of $\tilde{\Omega}_\text{HI}$, and we compare the analytic model with these observations in Fig. \ref{Figure: omegaHI}.
\cite{VillaescusaNavarro2018} simulate the behaviour of $\tilde{\Omega}_\text{HI}$ and find  good agreement with the observations.

\subsection{The 3D angular power spectrum}

\begin{figure}
\includegraphics[width=0.45\textwidth]{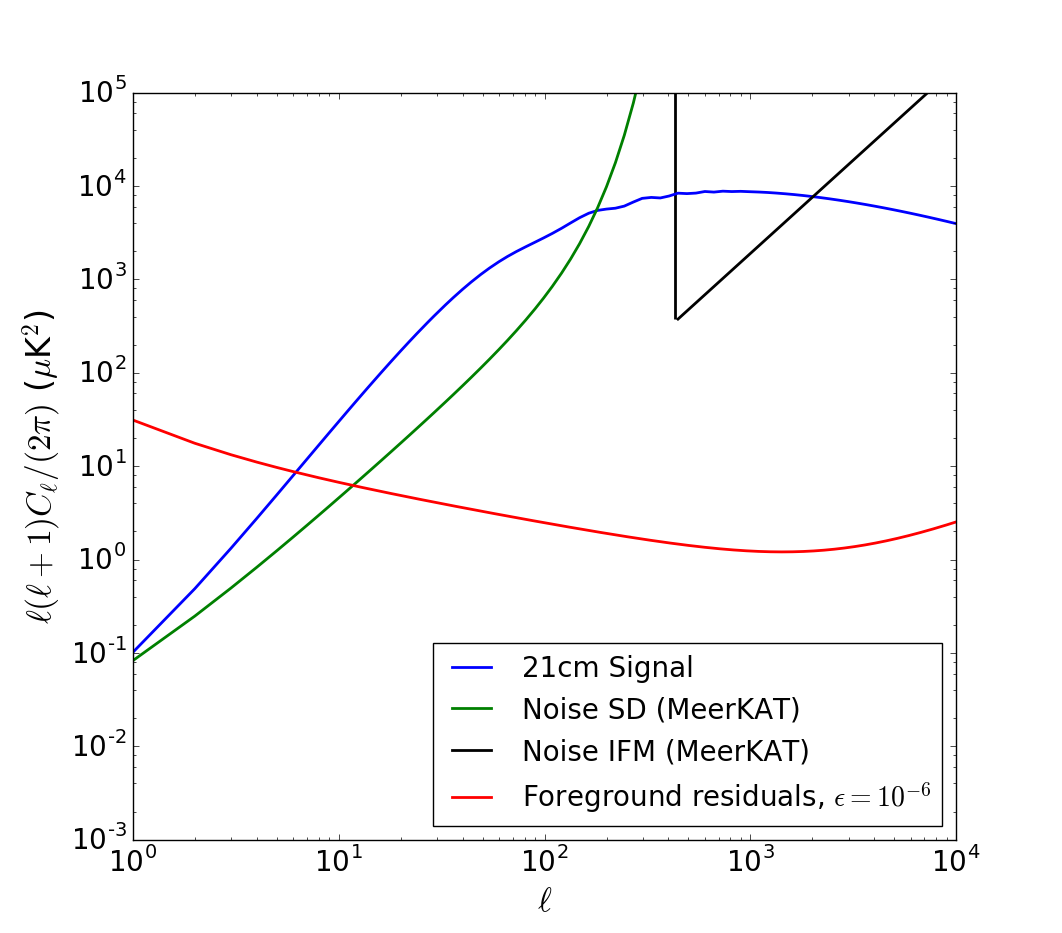}
\caption{Angular 21cm power spectrum, noise and foreground residuals at $z = 1$. We show the noise curves for MeerKAT operated in single-dish (SD, green) mode as well as in interferometer (IFM, black) mode. Foreground residuals are plotted for a removal efficiency of $\epsilon = 10^{-6}$.}\label{Figure: cls}
\end{figure}

Similarly to CMB experiments, fluctuations of the 21cm brightness temperature on the sky allow us to construct an angular power spectrum.
21cm experiments have a smooth frequency response around a central observed frequency $\nu$, which we model with a Gaussian window function $W_\nu(z)$, such that the observed brightness temperature fluctuation on the sky can be written as
\begin{equation}\label{eq: Window function relation}
\delta T_{\rm b}^{\text{obs}} (\hat{\boldsymbol{n}},\nu) =  \int dz W_\nu(z) \delta T_{\rm b}[r(z)\hat{\boldsymbol{n}},z].
\end{equation}
The quantity $\delta T_{\rm b}^{\text{obs}}$ thus denotes the observed temperature field projected onto the sky in a frequency bin labelled by $\nu$. 
As seen before, the brightness temperature fluctuations depend on the underlying HI density field.
At late times, most of the neutral hydrogen is located in self-shielded gas clouds inside galaxies, which means that the hydrogen density field is a biased tracer of the dark matter density field,
\begin{equation}\label{eq: biased tracer dTb}
\delta T_{\rm b}[r(z)\hat{\boldsymbol{n}},z] =  \delta\bar{T}_{\rm b}(z) \{1+b_\text{HI}(z) \delta[r(z)\hat{\boldsymbol{n}}, z]\}.
\end{equation}
In most of our analysis we are only concerned with the first order term as the monopole term is inaccessible through interferometry.
Similarly to \cite{Battye2013} and \cite{Bull2015} we assume the bias to be a constant at low-redshifts.
For our computations we fix $b_\text{HI} = 2$, which is consistent with DLA observations \citep{Font-Ribera2012, Hall2013}.
To first order in perturbation theory, the density fluctuations simply grow as a function of the growth factor,
\begin{equation}
\delta[r(z)\hat{\boldsymbol{n}}, z] = D_+(z)\delta(\boldsymbol{r}).
\end{equation}
We then Fourier transform the density fluctuations, and suppress the explicit $z$ dependence in our notation for simplicity, 
\begin{equation}
\delta(\boldsymbol{r}) = \int\frac{d^3\boldsymbol{k}}{(2\pi)^3} \tilde{\delta}(\boldsymbol{k}) e^{i\boldsymbol{k}\cdot\boldsymbol{r}},
\end{equation}
and subsequently expand the Fourier modes in spherical harmonics,
\begin{equation}\label{eq: exp}
e^{i\boldsymbol{k}\cdot\boldsymbol{r}} =4\pi\sum\limits_{\ell m}i^\ell j_\ell(kr)Y_{\ell m}(\hat{\boldsymbol{k}})Y^*_{\ell m}(\nhat).
\end{equation}
We find
\begin{equation}
\begin{aligned}
\delta T_{\rm b}^{\text{obs}} (\hat{\boldsymbol{n}},\nu) =  &4\pi\sum\limits_{\ell m}i^\ell\int dz W_\nu(z) \delta \bar{T}_{\rm b}(z) b_\text{HI}(z)D_+(z)\\
& \times \int \frac{d^3\boldsymbol{k}}{(2\pi)^3} \tilde{\delta}(\boldsymbol{k})j_\ell[kr(z)]Y_{\ell m}(\hat{\boldsymbol{k}})Y^*_{\ell m}(\nhat).
\end{aligned}
\end{equation}
Using the definition of the harmonic transform of the signal on the sky in terms of multipole moments $\ell$ and $m$,
\begin{equation}\label{eq: alm definition}
a_{\ell m}^\nu = \int d^2\nhat \delta T_{\rm b}^{\text{obs}} (\hat{\boldsymbol{n}},\nu) Y_{\ell m}(\nhat),
\end{equation} 
we can use the closure relation for spherical harmonics to obtain
\begin{equation}\label{eq: alm definition 2}
\begin{aligned}
a_{\ell m}^\nu =  &4\pi i^\ell\int dz W_\nu(z) \delta \bar{T}_{\rm b}(z) b_\text{HI}(z)D_+(z)\\
& \times\int \frac{d^3\boldsymbol{k}}{(2\pi)^3} \tilde{\delta}(\boldsymbol{k})j_\ell[kr(z)]Y_{\ell m}(\hat{\boldsymbol{k}}).
\end{aligned}
\end{equation}
Now, the angular 21cm power spectrum, $C_\ell$, is defined in terms of  ensemble average of two harmonic coefficients,
\begin{equation}\label{eq: Cl definition}
\left\langle a_{\ell m}^{\nu_1} a^{*\nu_2}_{\ell'm'}\right\rangle = \delta^{\rm K}_{\ell\ell'}\delta^{\rm K}_{mm'}C_\ell(\nu_1,\nu_2),
\end{equation}
where $\delta^{\rm K}$ denotes the Kronecker delta function and we assume statistical isotropy.
Combining equations \eqref{eq: alm definition 2} and \eqref{eq: Cl definition}, in conjunction with the Fourier space matter power spectrum relation, 
\begin{equation}
\left\langle\tilde{\delta}(\boldsymbol{k})\tilde{\delta}(\boldsymbol{k}')\right\rangle =  (2\pi)^3 \delta^{\rm D}(\boldsymbol{k} + \boldsymbol{k}') P(k),
\end{equation}
where $\delta^D$ is the Dirac delta function, we find the angular power spectrum to be
\begin{equation}
\begin{aligned}
C_\ell(\nu_1, \nu_2) = & \frac{2}{\pi}\int dz W_{\nu_1}(z) \delta \bar{T}_{\rm b}(z) b_\text{HI}(z)D_+(z)\\
&\int dz' W_{\nu_2}(z') \delta \bar{T}_{\rm b}(z') b_\text{HI}(z')D_+(z')\\
& \times\int dk k^2 P(k)j_\ell[kr(z)]j_\ell[kr(z')].
\end{aligned}
\end{equation}
For large $\ell$ we can use the Limber approximation (see \cite{Loverde2008}, equation \eqref{eq: limber}) such that the angular power spectrum becomes diagonal in frequency and reduces to
\begin{equation}\label{eq: final power spectrum}
C_\ell(\nu) = b_\text{HI}^2 \int dz \left[\frac{W_{\nu}(z) \delta \bar{T}_{\rm b}(z) D_+(z)}{r(z)}\right]^2 \frac{P\left[\frac{\ell+1/2}{r(z)}\right]}{|r'(z)|}.
\end{equation}
We compute the matter power spectrum, $P(k)$, using CAMB\footnote{Publicly available at: \url{https://camb.info/}.}, and  
our results for the 21cm angular power spectrum are illustrated in Fig. \ref{Figure: cls}, including our noise and foreground models described in section \ref{sec: instr}.

\section{Angular 21cm Bispectrum} \label{sec: bispectrum}

At low-redshifts ($z\sim 1$), targeted by upcoming IM experiments, the dark matter density field has become non-Gaussian mainly due to the non-linear gravitational collapse of structure.
As such, we expect the 21cm signal to contain a non-zero bispectrum.
The nature of the bispectrum provides a radical increase of observable modes as compared to the power spectrum and thus presents a promising probe for cosmology. 
Non-Gaussianity can be added to the signal through multiple channels, and here we focus on two main effects that contribute to the 21cm bispectrum. 
In addition to the non-Gaussianities due to structure formation, if the primordial density fluctuations are non-Gaussian, then that non-Gaussianity permeates through to late times as a contribution to the 21cm signal.
Furthermore, line of sight effects due to the gravitational distortion of light around massive objects and the accelerated expansion of the Universe, specifically via the integrated Sachs-Wolfe effect \citep{Sachs1967}, induce a non-Gaussian contribution to the signal. 
We can then write the total angular 21cm bispectrum as the sum of these contributing effects. 
Let $\alpha \equiv (\ell_1, \ell_2, \ell_3, m_1, m_2, m_3)$, then
\begin{equation}
B^\text{total}_\alpha = B_\alpha^\text{NLG} + B_\alpha^\text{LISW} + B_\alpha^\text{PNG}.
\end{equation} 

\subsection{Non-linear gravity bispectrum}
The bispectrum due to non-linear gravitational collapse of structure in the context of 21cm brightness temperature fluctuations can be calculated similarly to that in the context of galaxy surveys (See \cite{Fry1984} for details).
The brightness temperature fluctuations are sourced by the fluctuations in the HI field, which is a biased tracer of the DM field (see equation \eqref{eq: biased tracer dTb}). 
The bispectrum is then defined by the Fourier transform of the 3-point function,
\begin{equation}
\begin{aligned}
B^{21}(\boldsymbol{k}_1,\boldsymbol{k}_2,\boldsymbol{k}_3,z_1,z_2,z_3) &= \left\langle\delta\tilde{T}_{\rm b}(\boldsymbol{k}_1) \delta\tilde{T}_{\rm b}(\boldsymbol{k}_2) \delta\tilde{T}_{\rm b}(\boldsymbol{k}_3)\right\rangle \\
&=b_\text{HI}^3 \delta\bar{T}_{\rm b}(z_1)\delta\bar{T}_{\rm b}(z_2)\delta\bar{T}_{\rm b}(z_3) \\
& \times \left\langle \tilde{\delta}(\boldsymbol{k}_1,z) \tilde{\delta}(\boldsymbol{k}_2,z) \tilde{\delta}(\boldsymbol{k}_3,z)\right\rangle,
\end{aligned}
\end{equation}
where we assume a linear bias.
Expanding the density perturbations to second order and applying Wick's theorem, the lowest order contribution to the bispectrum is \citep{Pillepich2007} 
\begin{equation}
\begin{aligned}
B^{21}(\boldsymbol{k}_1,\boldsymbol{k}_2,\boldsymbol{k}_3,z_1,&z_2,z_3) = b_\text{HI}^3 2\mathcal{K}(\boldsymbol{k}_1,\boldsymbol{k}_2)D_+^2(z_1)\\
&D_+(z_2) D_+(z_3) \delta\bar{T}_{\rm b}(z_1)\delta\bar{T}_{\rm b}(z_2)  \\
& \delta\bar{T}_{\rm b}(z_3) P(\boldsymbol{k}_1)P(\boldsymbol{k}_2) + \text{cycl.},
\end{aligned}
\end{equation}
where we define 
\begin{equation}\label{eq: Kappa}
\mathcal{K}(\boldsymbol{k}_1,\boldsymbol{k}_2) \equiv A_0 + A_1\left(\frac{k_1}{k_2}+\frac{k_2}{k_1}\right)\cos\theta_{12} +A_2\cos^2\theta_{12},
\end{equation}
with $A_0 =5/7$, $A_1 = 1/2$, $A_2 = 2/7$, 
and $\theta_{12}$ denotes the angle between $\boldsymbol{k}_1$ and $\boldsymbol{k}_2$.

We can express the signal in harmonic space using equations \eqref{eq: Window function relation} and \eqref{eq: alm definition}.
Taking the ensemble average of three harmonic coefficients yields the angular bispectrum.
Using the methods developed in \cite{Verde2000} and \cite{Pillepich2007}, we compute the contribution to the angular 21cm bispectrum from the non-linear growth of structure to be
\begin{equation}\label{eq: ang bispectrum}
B_{\ell_1 \ell_2 \ell_3}^{\text{NLG}, m_1 m_2 m_3}(z) = B_{\ell_1 \ell_2 \ell_3}(z) \begin{pmatrix}
\ell_1 & \ell_2 & \ell_3 \\
m_1 & m_2 & m_3
\end{pmatrix},
\end{equation}
where the parentheses denote the Wigner-3J symbol, which ensures that the triangle condition is met, expresses isotropy, and is akin to the Kronecker delta in 3D space. 
The bispectrum is non-zero if and only if,
\begin{enumerate}
\item $-\ell_i \leq m_i \leq \ell_i$, for $i = 1,2,3$.
\item $m_1 + m_2 = - m_3$.
\item $|\ell_i - \ell_j|\leq \ell_k \leq \ell_i + \ell_j$, for all permutations of $(i,j,k) = (1,2,3)$.
\item $\ell_1+\ell_2+\ell_3$ is a non-zero integer unless $m_1 = m_2 = m_3 = 0$.
\end{enumerate} 
Further, we can write the bispectrum as a sum of cyclic terms,
\begin{equation}
B_{\ell_1 \ell_2 \ell_3}(z) = B_{12}(z) + B_{13}(z) + B_{23}(z),
\end{equation}
where
\begin{equation}\label{eq: NLG integral}
\begin{aligned}
B_{12}&(z) =  \frac{16}{\pi} i^{\ell_1+ \ell_2}\sqrt{\frac{(2\ell_1 + 1)(2\ell_2 + 1)(2\ell_3 + 1)}{(4 \pi)^3}} b_\text{HI}^3\\
&\times\int dz_1 dz_2 dz_3 dk_1 dk_2 k_1^2 k_2^2 P(k_1) P(k_2) \\
&\times W_\nu(z_1) W_\nu(z_2) W_\nu(z_3) D_+^2(z_1)D_+(z_2) D_+(z_3)\\ &\times\delta\bar{T}_{\rm b}(z_1) \delta\bar{T}_{\rm b}(z_2) \delta\bar{T}_{\rm b}(z_3)j_{\ell_1}[k_1 r(z_1)]j_{\ell_2}[k_2 r(z_2)] \\
&\times\sum\limits_{\ell \ell' \ell''} i^{\ell' + \ell''} (-1)^\ell \beta_\ell(k_1,k_2) (2\ell'+1) (2\ell'' + 1) \\
&\times j_{\ell'}[k_1 r(z_3)] j_{\ell''}[k_2 r(z_3)] 
\begin{Bmatrix}
\ell_1 & \ell_2 & \ell_3 \\
\ell'' & \ell' & \ell
\end{Bmatrix}
\\
&\times\begin{pmatrix}
\ell_1 & \ell' & \ell \\
0 & 0 & 0
\end{pmatrix}
\begin{pmatrix}
\ell_2 & \ell'' & \ell \\
0 & 0 & 0
\end{pmatrix}
\begin{pmatrix}
\ell_3 & \ell' & \ell'' \\
0 & 0 & 0
\end{pmatrix}.
\end{aligned}
\end{equation}
Here $\nu \equiv \nu(z)$, and we sum $\ell = 0, 1, 2$, $\ell' = \ell_1 - \ell, ... , \ell_1 + \ell$, and $\ell'' = \ell_2 - \ell, ... , \ell_2 + \ell$ and the braces denote the Wigner-6J symbol \citep[eg.][]{Sobelman1979}.
The $\beta_\ell(k_1,k_2) $ functions connect to \eqref{eq: Kappa} such that
\begin{equation}
\beta_0 = 2A_0 + \frac{2}{3}A_2\text{, }\beta_1 = 2 A_1\left(\frac{k_1}{k_2}+\frac{k_2}{k_1}\right) \text{, and }\beta_2 = \frac{4}{3} A_2.
\end{equation}
This expression can be simplified using the Limber approximation and we compute the three contributing $\ell$ terms separately in Appendix \ref{Appx: Limber},
\begin{subequations}
\begin{align}
B_{12}^{\ell = 0} = &b_\text{HI} A^{\ell = 0}_{\ell_1\ell_2\ell_3}  \int dz W_\nu(z)  \delta\bar{T}_{\rm b}(z)  D^2_+(z) \theta_{\ell_1}(z) \theta_{\ell_2}(z),\\
\begin{split}
B_{12}^{\ell=1} =& b_\text{HI} \sum\limits_{\ell'\ell''} A^{\ell = 1,\ell'\ell''}_{\ell_1\ell_2\ell_3}\int dz W_\nu(z) \delta\bar{T}_{\rm b}(z)D^2_{+}(z)\\
& \times  \left[\theta^1_{\ell_1\ell'}(z)\theta^{-1}_{\ell_2\ell''}(z)+\theta^{-1}_{\ell_1\ell'}(z)\theta^1_{\ell_2\ell''}(z)\right],\end{split}\\
\begin{split}
B_{12}^{\ell=2} = &b_\text{HI}\sum\limits_{\ell'\ell''} A^{\ell = 2,\ell'\ell''}_{\ell_1\ell_2\ell_3}\int dz W_\nu(z)  \delta\bar{T}_{\rm b}(z)D^2_{+}(z)\\
&\times \theta_{\ell_1\ell'}(z)\theta_{\ell_2\ell''}(z),
\end{split}
\end{align}
\end{subequations}
where the $A^\ell$ and the $\theta$-functions are defined in equations \eqref{eq: A0}, \eqref{eq: theta limber 0}, \eqref{eq: A}, \eqref{eq: theta l1},  \eqref{eq: A2}, and \eqref{eq: thetallpm2}.
In equation \eqref{eq: NLG integral} we have rederived the angular 21cm bispectrum due to non-linear gravitational collapse \citep[cf.][]{Pillepich2007} for our low-$z$ temperature model in equation \eqref{eq: temp model}.

\subsection{Lensing-ISW bispectrum}

The presence and evolution of the gravitational potential along the line of sight affects the 21cm radiation and imprints statistical information about the state of the matter distribution on the signal. 
Firstly, the photon paths are disturbed by the presence of gravitational wells, resulting in a weak lensing contribution to the signal.
The lensing potential, $\theta$, for a source at distance $r$ and at an angular position $\boldsymbol{\hat{n}}$ is a radial projection of the gravitational potential, $\Phi$, \citep{Bartelmann2001}.
In the Born approximation
\begin{equation}\label{eq: lensing potential}
\theta(r, \boldsymbol{\hat{n}}) = -\frac{2}{c^2} \int_0^r\, dr' \frac{S_k(r-r')}{S_k(r)S_k(r')}\Phi(r',\boldsymbol{\hat{n}}),
\end{equation}
where $S_k$ is determined by the curvature, and defined as 
\begin{equation}
S_k(r)=\left\{
                \begin{array}{ll}
                  \sqrt{k}^{-1}\sin{(r \sqrt{k})}, \hspace{22pt}k > 0,\\
                  r, \hspace{80pt}k = 0,\\
                  \sqrt{|k|}^{-1}\sinh{(r \sqrt{|k|})}, \hspace{4pt}k < 0.
                \end{array}
              \right.
\end{equation}
Observations of the weak lensing signal should be feasible by upcoming 21cm experiments and can help map the evolution of the growth function \citep{Pourtsidou2014}.

A second line of sight effect, sourced by the gravitational potential, affects the 21cm photons. 
Due to the accelerated expansion of the Universe at late times, potential wells evolve on timescales shorter than the crossing time for photons. 
Therefore, photons that enter the gravitational well obtain a boost in energy, which is higher than the required energy to leave the well due to the decay of the potential while crossing.
This results in an overall frequency gain which is additive along the photon's path. 
The frequency change due to this integrated Sachs-Wolfe (ISW) effect can be written as \citep{Nishizawa2014}  
\begin{equation}\label{eq: ISW2}
\frac{\Delta\nu}{\nu}(r,\boldsymbol{\hat{n}}) = \frac{2}{c^3}\int_0^r dr' \pd{\Phi(r',\boldsymbol{\hat{n}})}{t},
\end{equation}
where $t$ denotes the conformal time.

These line of sight effects perturb the apparent radial and angular position of the brightness temperature signal on the sky, $\delta T_{\rm b} = \delta T_{\rm b,0}(\boldsymbol{\hat{n}} + \nabla\theta, \nu +\Delta\nu)$, where $T_{\rm b,0}$ is the true, unperturbed signal, $\Delta\nu$ represents the frequency shift introduced by the ISW effect.
Expanding this signal to first order in the gravitational potential gives
\begin{equation}
\delta T_{\rm b} = \delta T_{\rm b,0} + \nabla\delta T_{\rm b,0}\cdot\nabla\theta + \nu \frac{d\delta T_{\rm b,0}}{d\nu} \frac{\Delta\nu}{\nu}.
\end{equation}
Considering a thin frequency shell, each term can be expanded in terms of multipole moments $\ell$ and $m$ on the sky via equation \eqref{eq: alm definition}.
Thus the total coefficients separate into contributions from the signal, the lensing gradient and the ISW frequency shift,
\begin{equation}\label{eq: Bispectrum total alm}
a_{\ell m}^\nu= a_{\ell m}^{0, \nu} + a^{\rm L,\nu}_{\ell m} + a^{\text{ISW}, \nu}_{\ell m}.
\end{equation}
The lensing coefficient is given by (Appendix \ref{Appx: lensing coefficient})
\begin{equation}
a_{\ell m}^{L,\nu} = \sum\limits_{\ell'm'\ell''m''} W_{\ell\ell'\ell''}^{mm'm''} a^{0,\nu*}_{\ell'm'} \theta^{\nu*}_{\ell''m''},
\end{equation}
where $W_{\ell\ell'\ell''}^{mm'm''}$ relates to the gaunt integral, $\mathcal{H}$ \citep[cf.][]{Verde2002}, via
\begin{equation}
W_{\ell\ell'\ell''}^{mm'm''} \equiv \frac{1}{2}(-1)^{m+m'+m''} L_{\ell\ell'\ell''} \mathcal{H}_{\ell\ell'\ell''}^{mm'm''},
\end{equation}
with
\begin{equation}
L_{\ell\ell'\ell''}\equiv -\ell(\ell+1) + \ell'(\ell'+1)+\ell''(\ell''+1).
\end{equation}

Taking the ensemble average of three harmonic coefficients, we note that the line of sight terms in equation \eqref{eq: Bispectrum total alm} are linear in the potential, such that linear terms vanish in the bispectrum and only second-order terms remain,
\begin{equation}
\begin{aligned}
\langle a^\nu_{\ell_1 m_1}&a^\nu_{\ell_2 m_2}a^\nu_{\ell_3 m_3}\rangle = \langle a_{\ell_1 m_1}^{0,\nu}a_{\ell_2 m_2}^{0,\nu}a_{\ell_3m_3}^{0,\nu}\rangle \\
&+ \sum\limits_{\substack{\ell'm'  \\ \ell'' m''}}W^{m_1m'm''}_{\ell_1\ell'\ell''} \left\langle a_{\ell'm'}^{*0,\nu}\theta_{\ell'' m''}^{*,\nu} a^{0,\nu}_{\ell_2 m_2} a_{\ell_3 m_3}^{\rm{ISW},\nu}\right\rangle \\
&+ {\rm perms.}
\end{aligned}
\end{equation}
The ISW and lensing effects are uncorrelated to the undisturbed signal, as the initial photon distribution from a distant source is not affected by any effects that distort this signal on the line of sight.
This allows us to separate the LISW contributions from the 21cm angular power spectrum in our expression for the bispectrum.
We find
\begin{equation}\label{eq:w }
\begin{aligned}
\langle a_{\ell_1 m_1}^\nu a_{\ell_2 m_2}^\nu a_{\ell_3 m_3}^\nu \rangle_\text{LISW}  &=  W_{\ell_1\ell_2\ell_3}^{m_1m_2m_3} C_{\ell_2}(\nu) Q_{\ell_3}(\nu) \\
&+ 5 \text{ perms.},
\end{aligned}
\end{equation}
where we have applied statistical isotropy to relate the 2-point statistics to the power spectra, 
\begin{equation}
\left\langle a_{\ell m}^{*0,\nu} a_{\ell' m'}^{0,\nu} \right\rangle= C_\ell(\nu) \delta^K_{\ell\ell'}\delta^K_{mm'},
\end{equation}
\begin{equation}\label{eq: Ql def}
\left\langle \theta_{\ell m}^{*\nu} a_{\ell' m'}^{\text{ISW},\nu} \right\rangle= Q_\ell(\nu) \delta^K_{\ell\ell'}\delta^K_{mm'}.
\end{equation}
The LISW power spectrum is given by \citep[cf.][]{Verde2002} (see Appendix \ref{Appx: LISW power})
\begin{equation}\label{eq: ql definition}
\begin{aligned}
Q_{\ell}(\nu) &= \frac{2\eta(z)}{c^4} \int_0^z\,dz'  \frac{S_k[r(z)-r(z')]}{S_k[r(z)]S_k[r(z')]r^2(z')} \\
&\times\left.\frac{\partial P_\Phi}{\partial z'}(k,z')\right|_{k=\ell/r(z')},
\end{aligned}
\end{equation}
where
\begin{equation}
\eta(z) = -(1+z)\frac{d\delta \bar{T}_{\rm b}}{dz}(z).
\end{equation}

The LISW bispectrum is a contamination to the primordial non-Gaussianity bispectrum observed on the CMB in both temperature and E-mode polarization \citep{Goldberg1999,Giovi2003,Lewis2011a}, and, if ignored, introduces a significant bias in the non-Gaussianity parameter $f_\text{NL}$ as measured by the \textit{Planck} mission using the skew-$C_\ell$ statistic \citep{Munshi2010}.
\cite{Planck2016NG} report a $2.8\sigma$ detection of the LISW bispectrum from temperature maps alone, which increases to a  $3\sigma$ detection including their polarization data.
We therefore compute the amplitude of the 21cm LISW bispectrum  and its effect as a contamination on the signal due to non-linear gravitational collapse.

\subsection{Primordial bispectrum}
Primordial non-Gaussianity in the density fluctuations are the most direct way to probe inflationary physics. 
Depending on the functional form of the inflaton field, PNG can be generated during inflation (see \cite{Bartolo2004} and \cite{Liguori2010} for extensive reviews).
The most accurate measurements of PNG to date \citep{Planck2016NG}, are consistent with perfectly Gaussian initial fluctuations.
However, with errors of order $\sigma_{ f_{\rm NL}} \sim 5-40$, depending on the triangle shape, the CMB cannot constrain the non-Gaussianity parameter on the $f_{\rm NL} \lesssim 1$ level, crucial for eliminating a variety of inflationary models such as models that include an early contraction phase \citep{Komatsu2009}.  
As any PNG would affect the distribution of dark matter in the Universe, and thus that of baryons, a contribution to the 21cm bispectrum is expected.
\cite{Pillepich2007} compute the angular bispectrum from PNG during the dark ages and compare it to the bispectrum from non-linear collapse. 
They find that the primordial bispectrum is $\sim 50$ times weaker than the gravitational bispectrum at large scales, but a cosmic variance limited experiment could produce competitive, $\sigma_{ f_{\rm NL}} \sim 1$, results.
Moreover, \cite{Munoz2015} study the 21cm bispectrum from PNG during the dark ages and find that 21cm observations can improve CMB constraints for PNG significantly due to the the high number of observable modes.
They predict that a cosmic variance limited experiment would be able to measure $f_{\rm NL}$ down to $\sigma_{ f_{\rm NL}} \sim 0.03$, and thus able to constrain single-field slow-roll inflation \citep{Maldacena2003,Acquaviva2003}.
Observations of PNG at lower redshifts rely on the scale dependence of the halo bias  and can achieve competitive constraints for the primordial non-Gaussianity parameter \citep{Mao2013, Aloisio2013, Li2017,Raccanelli2017,Karagiannis2018}.
The prospects of constraining PNG with the cosmic 21cm signal are thus promising. 

In this analysis, we focus on the information gain toward the cosmological parameters from the late-time 21cm bispectrum.
As the gravitational bispectrum dominates the bispectrum during the dark ages \citep{Pillepich2007}, the primordial bispectrum will remain sub-dominant at late times due to the progression of structure formation.
We will therefore ignore the PNG contribution to the bispectrum here.
\subsection{Bispectrum representation}\label{sec: Bispectrum Representation}
\begin{figure}
\includegraphics[width=0.48\textwidth]{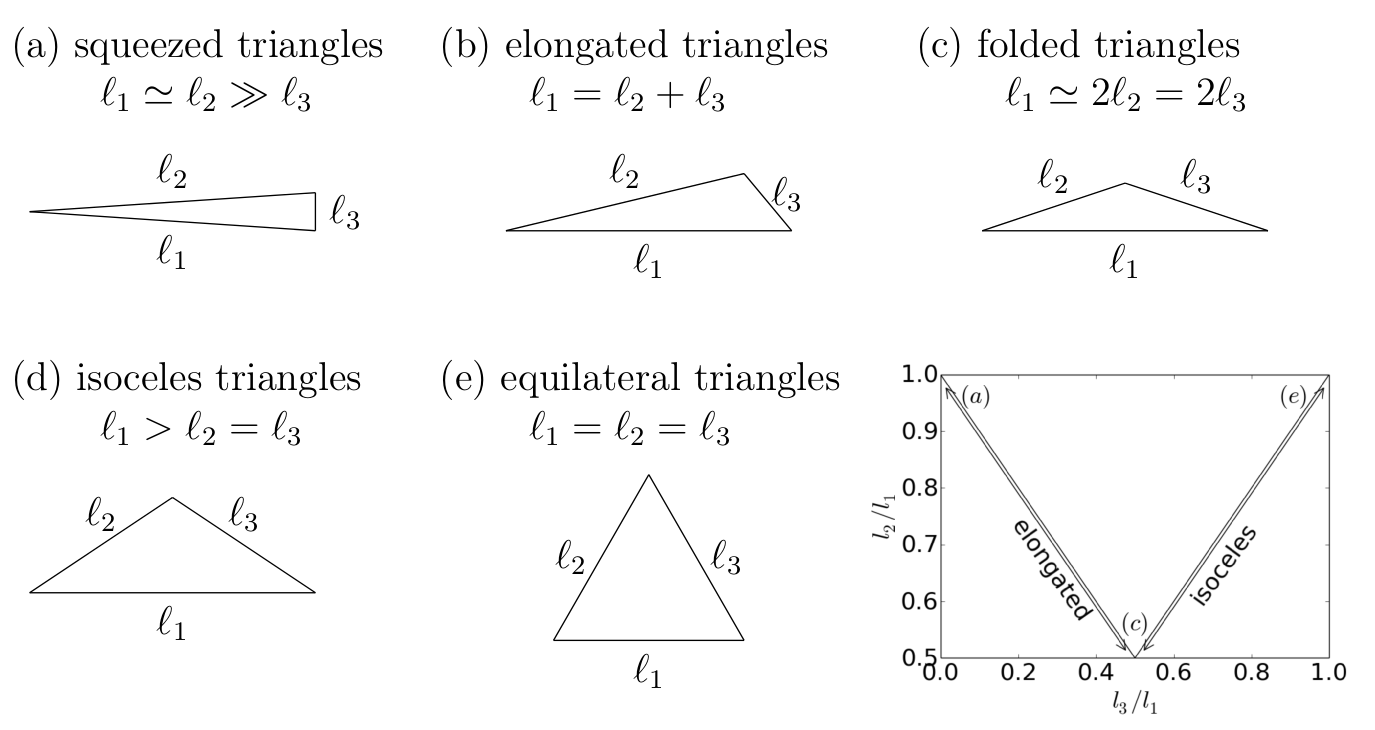}
\caption{We fix the order of the bispectrum modes to be $\ell_1 \geq \ell_2 \geq \ell_3$. 
Then, (a) - (e) show the relation between  triangle configurations and the bispectrum modes they represent.
When plotting the bispectrum as a function of the ratios of $\ell_2/\ell_1$ vs. $\ell_3/\ell_1$, different triangle configurations separate into different areas of the plot as shown in the lower right.
The bispectrum occupies a triangular shaped region which is due triangle condition obeyed by the bispectrum.
 }\label{Figure: triangles}
\end{figure}
\begin{figure*}
\includegraphics[width=0.80\textwidth]{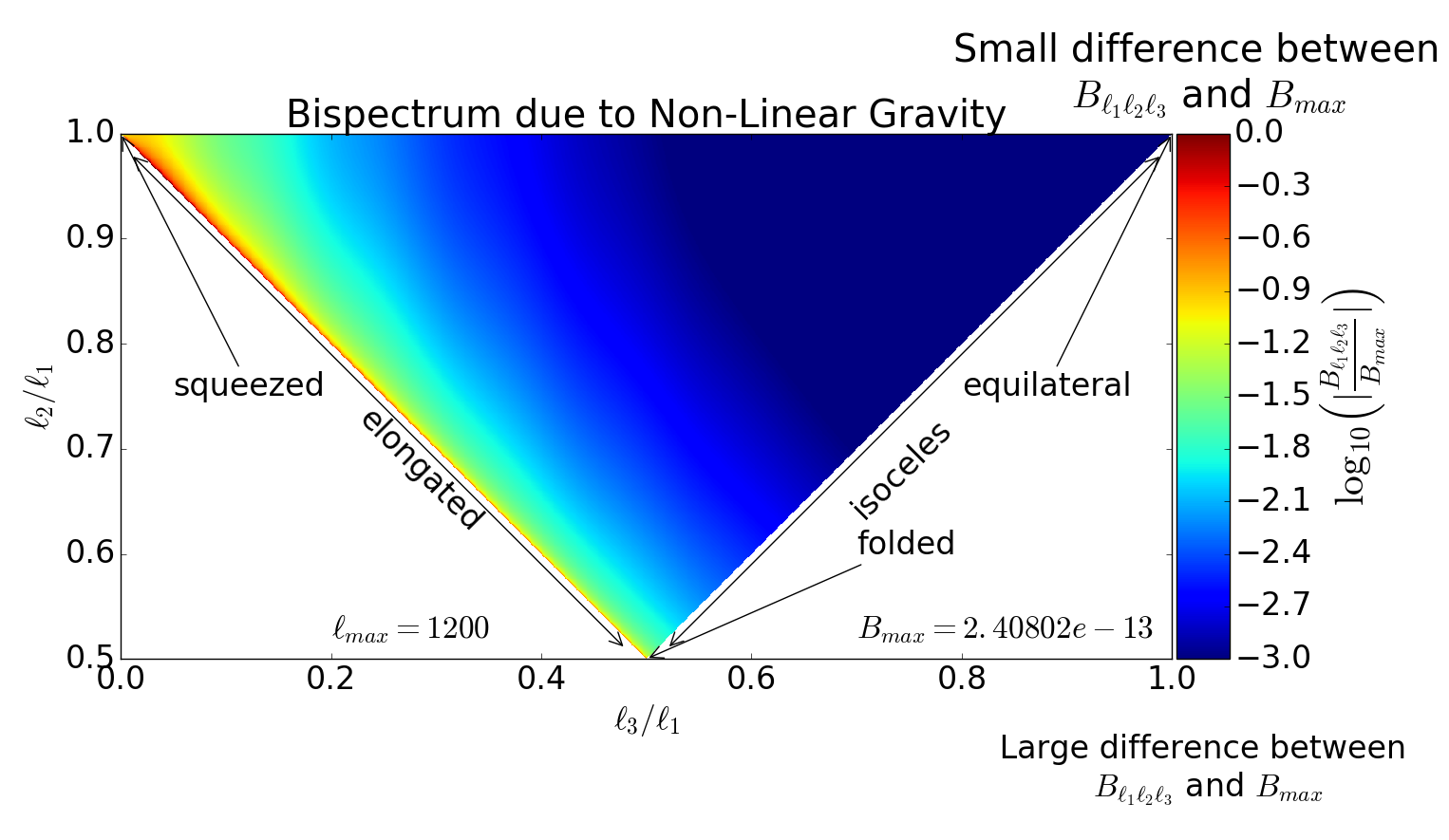}
\caption{We plot the amplitude of the angular bispectrum due to non-linear gravity collapse for $\ell_\text{max}$ at $z = 1$. 
The colour scale shows the order of magnitude difference in amplitude of the bispectrum relative to the triangle configuration with the largest bispectrum amplitude.
An elongated squeezed triangle configuration shows the largest amplitude with $B_\text{max} = 2.40802 \times 10^{-13}\text{mK}^3$.}\label{Figure: NLG Bispectrum}
\end{figure*}
\begin{figure*}
\includegraphics[width=0.80\textwidth]{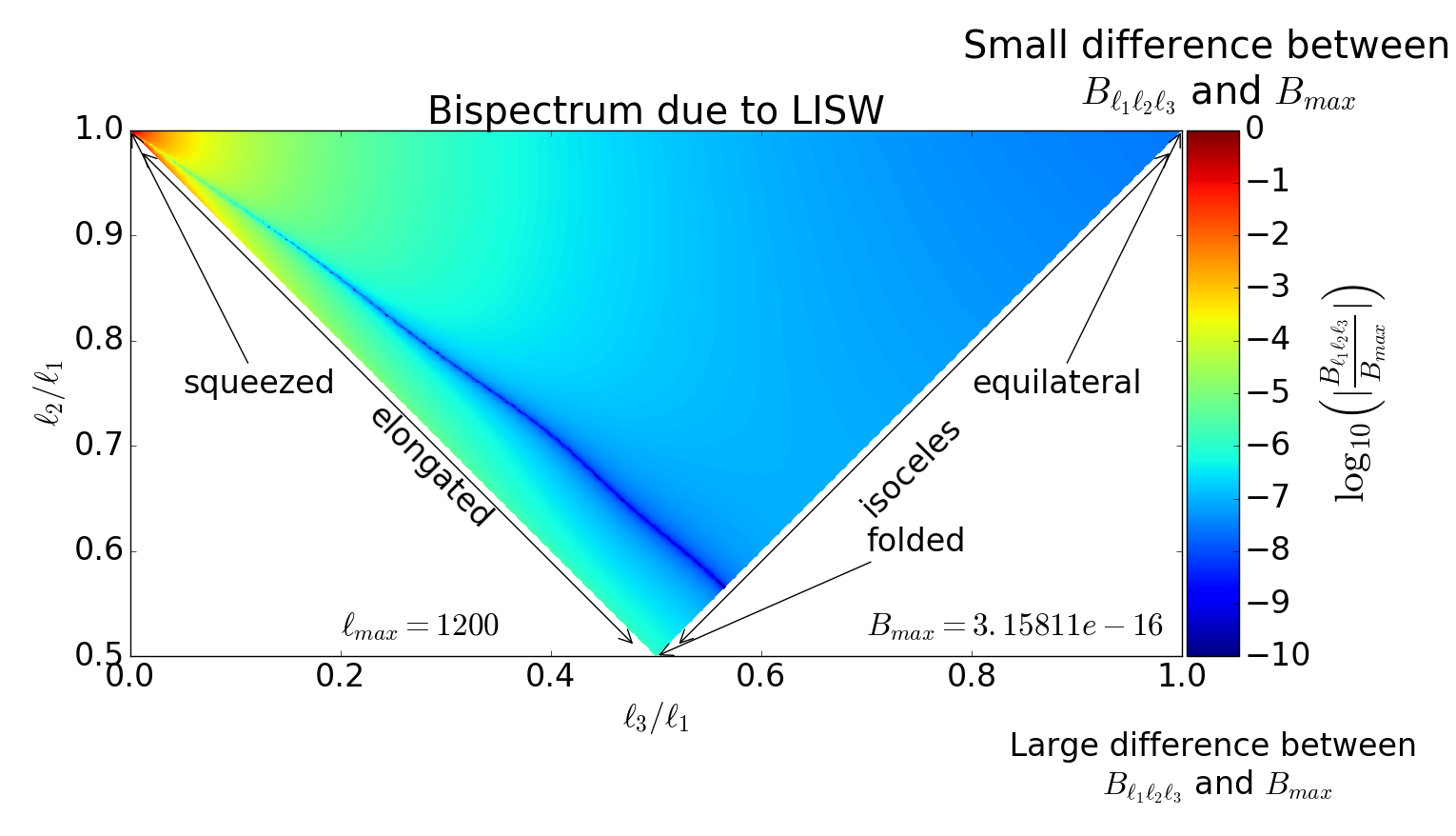}
\caption{We plot the amplitude of the angular LISW bispectrum for $\ell_\text{max}$ at $z = 1$. 
The colour scale shows the order of magnitude difference in amplitude of the bispectrum relative to the triangle configuration with the largest bispectrum amplitude.
The squeezed triangle configuration show the largest amplitude with $B_\text{max} = 3.15811 \times 10^{-16}\text{mK}^3$.
}\label{Figure: LISW Bispectrum}
\end{figure*}
\begin{figure}
\includegraphics[width=0.46\textwidth]{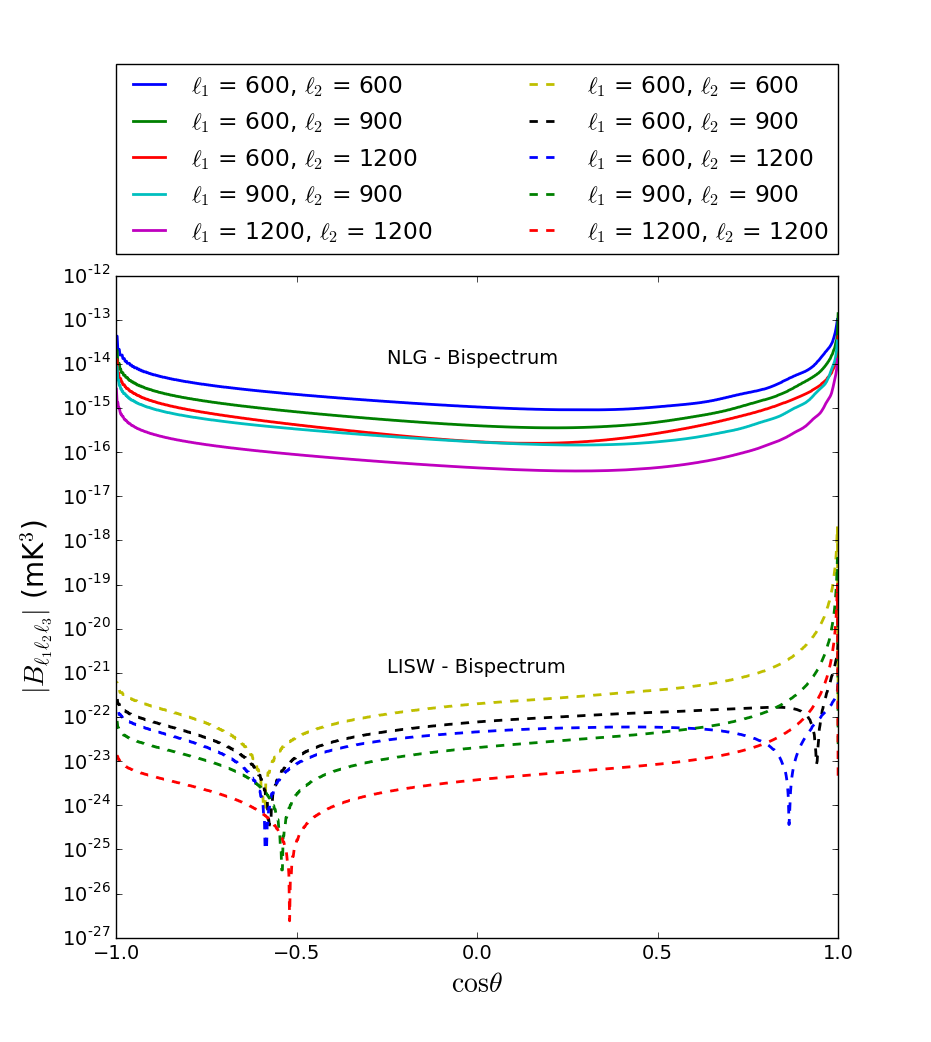}
\caption{Plot of the NLG (solid lines) and LISW (dashed lines) bispectrum as a function of the opening angle $\theta$ between two fixed triangle sides $\ell_1$ and $\ell_2$ at $z = 1$.}\label{Figure: Bispectrum2D}
\end{figure}
\begin{figure}
\includegraphics[width=0.46\textwidth]{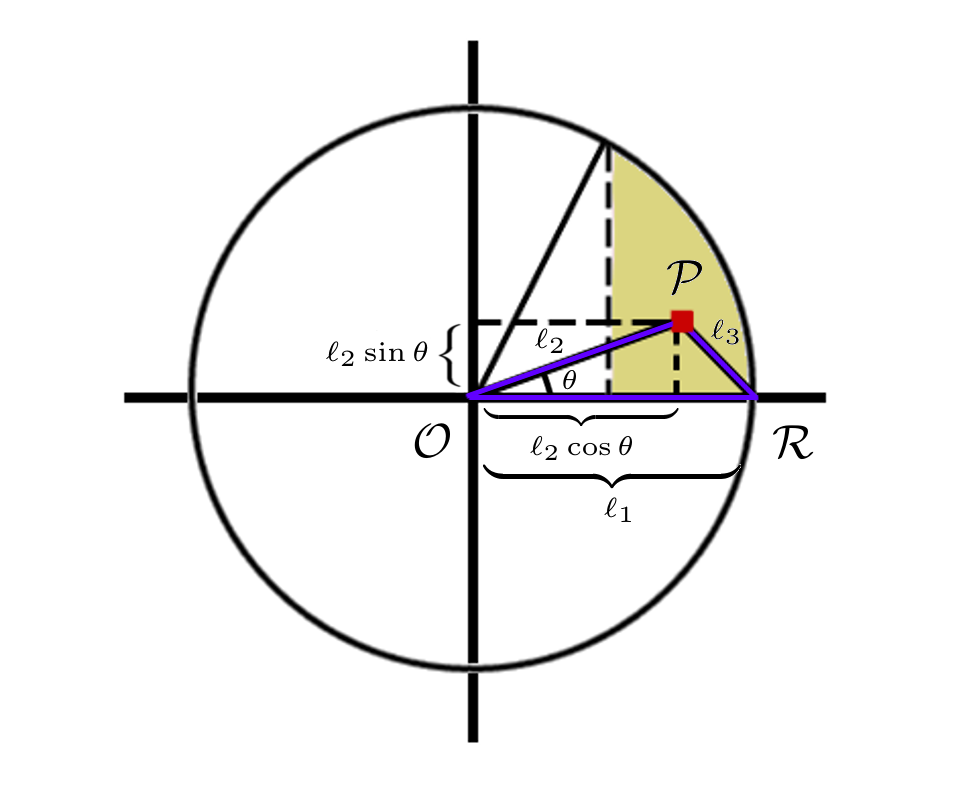}
\caption{This figure illustrates the interpretation of Fig. \ref{Figure: 2D bispectrum new}. 
All unique triangles with $\ell_1 \geq \ell_2 \geq \ell_3$ can be constructed when $\mathcal{P}$ lies within the shaded region, enclosed by the circle of radius $\ell_1$, the horizontal diameter of the circle, and the vertical line intersecting the circle at $\theta = \pi/3$.
Each pixel value in Fig. \ref{Figure: 2D bispectrum new}  corresponds to the bispectrum of the triangle configuration with $\ell_2 - \ell_3$ corner in the same location.  
}\label{Figure: bispectrum visual}
\end{figure}
\begin{figure*}
\includegraphics[width=0.8\textwidth]{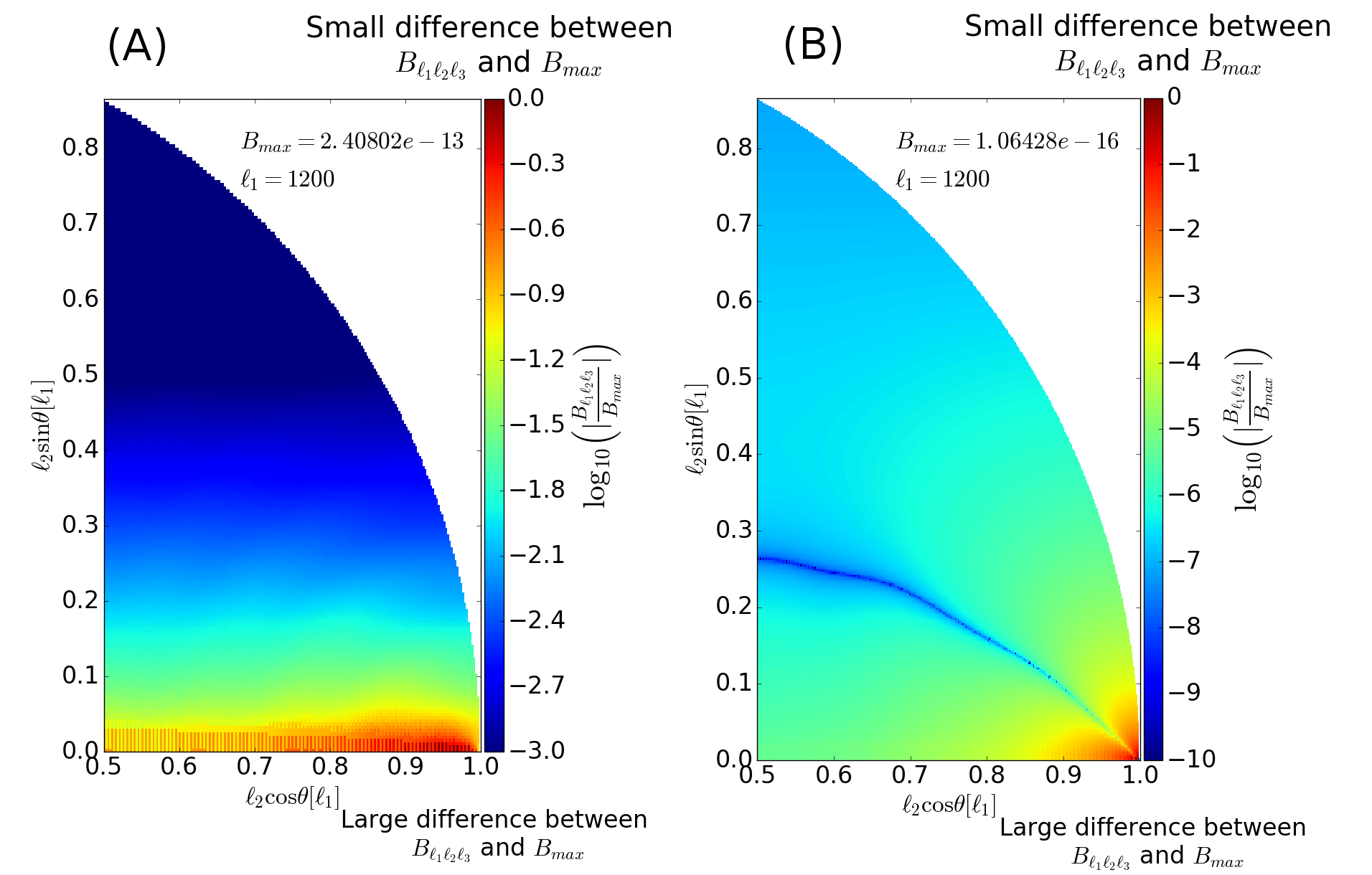}
\caption{We plot the amplitude of the angular (A) non-linear gravity and (B) LISW bispectrum for $\ell_\text{max} = 1200$ at $z = 1$. 
The $x$ and $y$ axes are in units of $\ell_1$.
The colour scale shows the order of magnitude difference in amplitude of the bispectrum relative to the triangle configuration with the largest bispectrum amplitude.
For (A), the non-linear gravity bispectrum, the elongated and squeezed triangle configurations show the largest amplitude with $B_\text{max} = 2.40802 \times 10^{-13}\text{mK}^3$.
For (B), the LISW bispectrum, the squeezed triangle configuration shows the largest amplitude with $B_\text{max} = 1.06428 \times 10^{-16}\text{mK}^3$.}\label{Figure: 2D bispectrum new}
\end{figure*}

The bispectrum can be represented geometrically as a correlation of the signal from the corners of a triangle where the length of the sides is related to the wavenumber of the bispectrum. 
For our computation of the bispectrum we use the triangular representation of \cite{Jeong2009} shown in Fig. \ref{Figure: triangles}. 
We set $\ell_1 \geq \ell_2 \geq \ell_3$, and fix $\ell_1$, while varying $\ell_2$ and $\ell_3$.
Plotting the ratios to the largest $\ell$-mode against each other results in a triangular plot where squeezed bispectrum configurations occupy the upper left corner, equilateral configurations occupy the upper right corner, folded triangles are in the triangle peak, and elongated and isoceles triangles occupy the sides of the triangle.
We show the relative amplitudes of our bispectrum calculations for $\ell = 1200$ at $z = 1$ for the NLG and the LISW bispectrum in Fig. \ref{Figure: NLG Bispectrum} and \ref{Figure: LISW Bispectrum} respectively. 
In order to visualize the overall trend of the bispectrum as a function of triangle configuration, we interpolate the bispectrum between neighbouring pixels as the statistical isotropy of the signal requires the sum of modes to be even and thus renders every other pixel zero.

We see for both cases that most of the signal is coming from squeezed or quasi-squeezed triangle configurations.
The non-linear gravity bispectrum shows a large contribution from elongated triangles and three orders of magnitude lower contributions from equilateral triangle configurations. 
Similarly, the LISW bispectrum experiences almost no contributions from equilateral triangles. 
The dark blue stripe in Fig. \ref{Figure: LISW Bispectrum} is due to a sign flip of the bispectrum and the bispectrum approaches zero for triangle configurations close to the feature.

An alternative representation with a single  degree of freedom is presented in \cite{Majumdar2018}, where the bispectrum is plotted for two fixed side lengths as a function of the opening angle of the triangle.
Fig. \ref{Figure: Bispectrum2D} shows our results of NLG and LISW bispectra as a function of the opening angle for two fixed side lengths and illustrates the large amplitude difference of $\sim 7$ orders of magnitude between the non-linear gravity and LISW bispectrum for these modes.

Here, we also propose a new representation for the bispectrum which gives a direct visual connection to the triangle configuration at each point.
The sketch in Fig. \ref{Figure: bispectrum visual} shows the interpretation of this representation.
Fig. \ref{Figure: 2D bispectrum new} shows our `sail' plots of the bispectrum for NLG and LISW which contain the bispectrum values for all unique triangle shapes. 
We fix the longest side of a triangle to be the horizontal radius of a circle of length $\ell_1$, $\overline{\mathcal{OR}}$.
When labelling the second largest side $\ell_2$, we require $\ell_2\cos\theta \in [\frac{1}{2} \ell_1, \ell_1]$ and $\ell_2\sin\theta \in [0, \frac{\sqrt{3}}{2}\ell_1]$ to construct all possible unique triangles, as any others are obtained through rotation and relabelling of the sides.
Now, for each point $\mathcal{P}$ in the shaded region, we compute the bispectrum of the corresponding triangle configuration and show the result as a colour scale at that point. 
We thus produce a colour map, where the $x-y$ coordinates are identical to the coordinates of the point $\mathcal{P}$ of the corresponding triangle, allowing for a direct and natural interpretation of the map.

Fig. \ref{Figure: 2D bispectrum new} panel (A) shows the same behaviour as Fig. \ref{Figure: NLG Bispectrum}, where the largest bispectrum is obtained by squeezed triangles, close to the $x$-axis, and the lowest in the equilateral limit. 
Further, Fig. \ref{Figure: 2D bispectrum new} panel (B) can be interpreted in the same way and compared to Fig. \ref{Figure: LISW Bispectrum}.
Whereas the triangle plots allow the bispectrum for a given triplet of modes $(\ell_1,\ell_2,\ell_3)$ to be read directly, connecting regions of the plot with particular triangle shapes can be cumbersome. 
On the one hand, the triangle shape corresponding to any given pixel value in our `sail' plots can directly be read off by constructing a triangle according to Fig. \ref{Figure: bispectrum visual}.
On the other hand, reading the corresponding $(\ell_1,\ell_2,\ell_3)$ triplet may not be straightforward, as $\ell_2 =  \sqrt{x^2 + y^2}$, and $\ell_3 = \sqrt{\ell^2_1+x^2 + y^2-2\ell_1 x}$.

For both triangle plots and `sail' plots the full bispectrum information is only obtained when stacking the plots for all different values of $\ell_1$. 
We have included figures at $\ell_1 = 1200$ as an example of the value of the bispectrum.

\section{Instrument and Foreground Assumptions} \label{sec: instr}
In this section we explore instruments and foregrounds which will both limit the detectability and sensitivity of the quantities derived up to this point. 
\subsection{Instruments}
We examine three different experiments in this analysis: CHIME, MeerKAT and SKA.

\textit{CHIME}\footnote{see \url{https://chime-experiment.ca/}}\textit{:} 
The Canadian Hydrogen Intensity Mapping Experiment (CHIME), based in British Columbia, is an interferometer consisting of four 100 x 20 metre semi-cylinders equipped with radio receivers sensitive to 400MHz - 800MHz ($z \sim 0.8 - 2.5$). 
This experiment is a dedicated low-redshift 21cm intensity mapping experiment targeting BAO scales, with applications in FRB detection and pulsar monitoring.
We select this telescope as a currently operational intensity mapping experiment, with the potential for late-time 21cm signal detection.
 
\textit{MeerKAT}\footnote{see \url{http://www.ska.ac.za/science-engineering/meerkat/}}\textit{:} MeerKAT is an array of sixty-four 13.5 metre dishes located in the Karoo desert in South Africa.
The dishes are equipped with three separate receivers, with the low-frequency band going from 580 MHz to 1015 MHz ($z \sim 0.4 - 1.4$).
This SKA precursor will eventually be fully integrated into SKA-MID.
We select MeerKAT as it is a near-future 21cm experiment with the potential to do low-redshift intensity mapping and is a precursor of SKA-mid.

\textit{SKA-MID}\footnote{see \url{https://www.skatelescope.org/mfaa/}}\textit{:} The SKA Mid-Frequency Aperture Array is the South African part of the multi-purpose Square Kilometre Array and will consist of 190 15 metre dishes, at the MeerKAT location in the Karoo desert.
SKA-Mid will be able to perform intensity mapping in both single-dish (autocorrelation) and interferometer mode.
We focus on the 350 MHz - 1050 MHz ($z \sim 0.35 - 3$) range which SKA-MID will be operating at.
SKA-mid is selected to illustrate the degree to which observations may constrain cosmology over the next decade.  

\subsection{Instrumental noise}
CHIME, MeerKAT and SKA-MID can all be operated in an interferometric mode, where the noise power spectrum can be modelled as \citep{Zaldarriaga2004, Pourtsidou2014}
\begin{equation}
C_\ell^{\rm N} = \frac{T_\text{sys}^2(2\pi)^2}{\Delta\nu t_{\rm o} f_\text{cover}^2\ell^2_\text{max}},
\end{equation}
where $T_\text{sys}$ is the system temperature of the dishes in the array,  $t_o$ is the total observing time and $\ell_\text{max}=2\pi D_\text{tel} / \lambda$ determines the largest multipole moment accessible by an array with diameter $D_\text{tel}$ at an observed wavelength $\lambda$. 
The covering fraction, $f_\text{cover}$, is the ratio of the collecting area, $A_\text{coll}$, to the physical area covered by the array, st. $f_\text{cover} = A_\text{coll} / [\pi (D_\text{tel}/2)^2]$.
The system temperature is given as the sum of the angle-averaged sky temperature and the temperature of the antenna, $T_\text{sys} = T_\text{ant} + T_\text{sky}$.
At the frequencies considered in our analysis, the system temperature is dominated by $T_\text{ant}$ at $\sim 30 - 50\text{K}$, such that we assume $T_\text{sys} \approx T_\text{ant}$.
Interferometers cannot resolve scales larger than those set via the minimal baseline, which we model as a sharp noise increase at $\ell < \ell_\text{min} = 2\pi D_\text{min} / \lambda$.

MeerKAT and SKA-MID can also be operated in single-dish mode, for which the thermal noise per beam is given via \citep{Olivari2018}
\begin{equation}
\sigma_{\rm t} = \frac{T_\text{sys}}{\sqrt{t_\text{pix}\Delta\nu}},
\end{equation}
where $T_\text{sys}$ is the system temperature of the dishes, $\Delta\nu$ is the frequency binwidth and $t_\text{pix}$ is the integration time per beam.
The integration time per beam is obtained by distributing the total integration time $t_{\rm o}$, across $N_{\rm d}$ dishes, 
\begin{equation}
t_\text{pix} = N_{\rm d} t_{\rm o} \frac{\theta^2_\text{FWHM}}{S_\text{area}},
\end{equation}
where $\theta^2_\text{FWHM} = \pi^2 / \ell_\text{max}^2 $ is the beam area, which sets the smallest scale that can be observed by each dish as $\ell_\text{max} = 2\pi D_\text{dish} / \lambda$, where we use the diameter of the dish,  $D_\text{dish}$, as opposed to the diameter of the array in the interferometer case, and $S_\text{area}$ denotes the survey area.
Further, the signal is suppressed by the beam when angular wavenumbers exceed the resolution of the instrument.
We model this beam suppression as an exponential increase in the noise as a function of $\ell$,
\begin{equation}
C^{\rm N}_\ell = e^{\sigma^2 \ell^2} C_\ell^{\rm N,\text{thermal}}, 
\end{equation}
with $\sigma^2 = \theta^2_\text{FWHM}/8\ln 2$, setting the scale of signal suppression.
The full noise power spectrum for a single-dish array is then given by \citep[eg.][]{Dodelson2003}
\begin{equation}
C_\ell^{\rm N} = \sigma_{\rm t}^2 \theta^2_\text{FWHM} e^{\frac{\theta^2_\text{FWHM} \ell^2}{8\ln 2}}.
\end{equation}
The noise parameters for each experiment are listed in table \ref{table: noise params}.
Experiments will likely bin their observations into bins with $\Delta\nu \leq 1\text{MHz}$ \citep{Pourtsidou2016}, however decreasing the window width for bispectrum observations increases the computation run-time to levels of impracticality, such that we take a conservative bin width of $\Delta\nu = 10 \text{MHz}$. 
As the target emission is sourced by the discrete galaxy population, a shot noise contribution is expected.
\cite{Chang2008} find the shot noise contribution in the case of post-EoR intensity mapping observations to be negligible, and we ignore this term here.
One of the biggest challenges for radio observations of the 21cm signal will be spectral calibration as a means to remove large foreground contaminations.
In this context, recently the closure phase techniques of \cite{Thyagarajan2018} and \cite{Carilli2018} have used the bispectrum phase as a calibration tool for redundant baseline arrays, but highlight its usefulness for HI intensity mapping experiments.
We note that a connection between the closure phase and the angular bispectrum may be of interest, however we leave an analysis of this connection for future work.

\begin{table}
\begin{center}
\caption{Experimental noise parameters for CHIME, MeerKAT and SKA-Mid. We include parameters for both interferometry and single-dish noise models.}
 \begin{tabular}{c @{\hskip 0.15cm} c @{\hskip 0.15cm} c@{\hskip 0.15cm} c } 
 \hline
 Parameter & CHIME & MeerKAT & SKA-Mid \\ [0.5ex] 
 \hline\hline
 $\nu_\text{min}$  & 400 MHz  & 580 MHz  & 350 MHz\\
 \hline
$\nu_\text{max}$ & 800 MHz & 1020 MHz & 1050 MHz\\
 \hline 
 $T_\text{ant}$ & 50 K & 29 K & 28 K \\
 \hline  
  $D_\text{tel}$ & 100 m & 800 m & 1 km \\
 \hline  
 $D_\text{min}$ & 20 m & 29 m & 34 m \\
 \hline 
  $A_\text{coll}$ & 8000 m$^2$ & 9000 m$^2$ & 33000 m$^2$ \\
 \hline  
  $N_\text{d}$ & 4 &  64 & 190 \\
 \hline  
  $D_\text{dish}$ & 20 m & 13.5 m & 15 m \\
 \hline  
 $S_\text{area}$ & 25000 deg$^2$ & 25000 deg$^2$ & 25000 deg$^2$ \\
 \hline  
  $t_{\rm o}$ & $10^4$ hours & $10^4$ hours & $10^4$ hours \\
 \hline  
  $\Delta\nu$ & 10 MHz & 10 MHz & 10 MHz \\
 \hline  
\end{tabular}
\label{table: noise params}
\end{center}
\end{table}

\subsection{Foregrounds}

Cosmological 21cm observations suffer from large foreground contaminations from both galactic and extragalactic sources.
Successful detections of the signal hinge strongly on the accurate modelling and removal of these contaminations  which can be 4 to 6 orders of magnitude larger than the signal
\citep{Liu2009,Alonso2014}.
Here, we model four contaminating foreground sources \citep{Santos2005}: extragalactic point sources, extragalactic free-free emission, galactic synchrotron emission and galactic free-free emission.
Due to the smooth frequency variation of these foregrounds, a variety of foreground removal strategies have been proposed \citep{Oh2003,Barkana2005,Wolz2014,Alonso2015}.
All of these however leave some degree of residual amplitude on the signal. 
These residuals are often modelled as a power law \citep{Bull2015}, and we will adopt this model here.
Our foreground residuals are modelled as
\begin{equation}
C_\ell^{\rm FG}(\nu)  = \epsilon^2 \sum\limits_X A_X \left(\frac{\ell_{\rm f}}{\ell}\right)^{n_X} \left(\frac{\nu_{\rm f}}{\nu}\right)^{m_X},
\end{equation}
where the sum is taken over all contributing sources $X$. 
The power law coefficients $n_X$ and $m_X$, as well as the amplitudes associated with each foreground are listed in table \ref{table: foregrounds}. 
Similar to \cite{Bull2015}, we multiply our foreground model with a removal efficiency coefficient $\epsilon$. 
Then, $\epsilon = 1$, if no foreground removal has been applied.
In our analysis we adopt an optimistic value of $\epsilon = 10^{-6}$ as we aim to determine the capabilities of a noise-limited detection of the bispectrum. 
When this value is raised to higher than $\sim 10^{-5}$, we observe significant deterioration of the constraints.
Fig. \ref{Figure: cls} then shows an example comparison of the foreground residuals removed with $\epsilon = 10^{-6}$ to the noise models of MeerKAT in single-dish and interferometer mode and the power spectrum model at $z = 1$.
We would like to stress however that this foreground model is simplistic and a thorough analysis of the significance of foregrounds in the context of observations of higher order statistics of the diffuse 21cm signal would be beneficial.

\begin{table}
\begin{center}
\caption{Foreground model parameters \citep{Santos2005}.}
 \begin{tabular}{c @{\hskip 0.15cm} c @{\hskip 0.15cm} c@{\hskip 0.15cm} c} 
 \hline
 Foreground source & $A_X [\text{mK}]^2$ & $n_X$ & $m_X$ \\ [0.5ex] 
 \hline\hline
Extragalactic point sources & 57 & 1.1 & 2.07 \\
 \hline
Extragalactic free-free & 0.014 & 1.0 & 2.1 \\
 \hline 
Galactic synchrotron & 700 & 2.4 & 2.8 \\
 \hline  
Galactic free-free & 0.088 & 3 & 2.15 \\
  \hline
  \hline

\end{tabular}
\label{table: foregrounds}
\end{center}
\end{table}

\section{Fisher Analysis}\label{sec: stats}

\subsection{The Fisher matrix}

The Fisher information matrix \citep{Fisher1935,Tegmark1997,Hobson2010} is a powerful tool which allows us to estimate the minimum error one can expect from an upcoming experiment by assuming that the likelihood assumes a multivariate Gaussian form in the model parameters. 
By Taylor expanding the log-likelihood around its maximum-likelihood value, one can define the Fisher matrix as
\begin{equation}
\boldsymbol{F}_{ij} \equiv \left\langle\frac{\partial^2 \mathcal{L}}{\partial\theta_i \partial\theta_j}\right\rangle,
\end{equation}
where $\mathcal{L} \equiv - \ln L$, the negative log-likelihood.
The Cramer-Rao inequality \citep[e.g.][]{Heavens2009, Hobson2010} then gives a lower bound for the errors one is expected to attain. 
When marginalizing over all other parameters in the analysis, the expected error on parameter $i$ is given by
\begin{equation}
\sigma_i \geq \sqrt{(\boldsymbol{F}^{-1})_{ii}},
\end{equation}
which reduces the problem of predicting the minimum errors for an experiment to computing the Fisher matrix and inverting it.
\cite{Tegmark1997} report the Fisher matrix for Gaussian data as
\begin{equation}
\boldsymbol{F}_{ij} = \frac{1}{2}\text{Tr}(\boldsymbol{A}_i\boldsymbol{A}_j + \boldsymbol{C}^{-1} \boldsymbol{M}_{ij}),
\end{equation}
where $\boldsymbol{C}$ denotes the covariance matrix, $\boldsymbol{A}_i \equiv \boldsymbol{C}^{-1} \boldsymbol{C}_{,i}$, $\boldsymbol{M}_{ij} \equiv \boldsymbol{\mu}_{,i}\boldsymbol{\mu}^T_{,j} + \boldsymbol{\mu}_{,j}\boldsymbol{\mu}^T_{,i}$, and $\boldsymbol{\mu} \equiv \left\langle\boldsymbol{x}\right\rangle$, where $\boldsymbol{x}$ denotes the data vector.
We use the standard comma notation to signify derivatives with respect to the parameter, $\boldsymbol{C}_{,i} \equiv \partial\boldsymbol{C}/\partial\theta_i$.

For power spectrum forecasts, the data vector is taken to be the angular coefficient observed at some frequency $\nu$, 
\begin{equation}
\boldsymbol{x}_{\ell m}^{\nu} = a_{\ell m}^{\nu}.
\end{equation} 
As $\mu = \left\langle a_{\ell m}^{\nu}\right\rangle = 0$, the second term in the trace vanishes and
 \begin{equation}\label{eq: Fisher power spectrum}
 \begin{aligned}
\boldsymbol{F}_{ij} &= \frac{1}{2}\text{Tr}(\boldsymbol{C}^{-1} \boldsymbol{C}_{,i}\boldsymbol{C}^{-1} \boldsymbol{C}_{,j}) \\
&= f_\text{sky}\sum\limits_\nu\sum\limits_\ell (2\ell+1) \frac{C_{\ell,i}^\nu C_{\ell,j}^\nu}{\left(C_\ell^{\nu,\text{tot}}\right)^2},
\end{aligned}
\end{equation}
where we have summed over all $m$ indices, and introduced a sky covering fraction $f_\text{sky}=0.5$ which effectively decreases the information gain by half and accounts for the correlation of nearby modes by the sky mask.
We assume that the signal, noise and foreground residuals are all uncorrelated to each other, thus we find that
\begin{equation}
C_\ell^{\nu,\text{tot}} = C_\ell^{\nu, \rm S} + C_\ell^{\nu, \rm N} + C_\ell^{\nu, \rm FG}.
\end{equation}

For the bispectrum analysis, the first term in the trace is small, due to the large number of triangles contributing to $C^{-1}$, and the Fisher matrix thus depends on the derivatives of the data with respect to the parameters only.
To ensure that the data vector for this analysis is Gaussian distributed, we use a weighted average of the angular bispectrum as the data 
\begin{equation}
\boldsymbol{x}_{\ell_1\ell_2\ell_3}^{\nu} = \sum\limits_{m_1 m_2 m_3} a^{\nu}_{\ell_1 m_1} a^{\nu}_{\ell_2 m_2} a^{\nu}_{\ell_3 m_3} w_{m_1 m_2 m_3}^{\ell_1 \ell_2 \ell_3},
\end{equation}
where it is easy to show that for an unbiased, minimum-variance estimator of the bispectrum, the weighting function is the Wigner-3J symbol, 
\begin{equation}
w_{m_1 m_2 m_3}^{\ell_1 \ell_2 \ell_3} = \begin{pmatrix}
\ell_1 & \ell_2 & \ell_3 \\
m_1 & m_2 & m_3
\end{pmatrix}.
\end{equation}
The above data vector can be shown to be Gaussian distributed for large $\ell$, by grouping elements of the sum such that each group shares the same value of the Wigner-3J symbol.
The central limit theorem can then be applied as the elements within each sum are independent and identically distributed random variables, making each group a Gaussian random variable itself.
Finally, the sum of all groups is a sum of Gaussian random variables and itself Gaussian distributed.

For computational ease, we assume that our bispectrum is uncorrelated between different frequency bins, such that we observe the bispectrum from a single frequency bin centred at $\nu$ only.
The total Fisher matrix is thus the sum of the contributions from all frequency bins and all contributing modes $\lambda \equiv (\ell_1, \ell_2, \ell_3)$, which obey the triangle conditions,
\begin{equation}
\boldsymbol{F}_{ij} = \sum\limits_\nu \sum\limits_{\lambda} \boldsymbol{F}_{ij}^{\nu, \lambda}.
\end{equation}
Then, applying Wick's theorem to evaluate the covariance matrix,
\begin{equation}
\boldsymbol{C} = \left\langle(\boldsymbol{x} - \boldsymbol{\mu})(\boldsymbol{x} - \boldsymbol{\mu})^t\right\rangle,
\end{equation}
after applying the sum \eqref{eq: sum W3J squared over m} in computing $\left\langle x\right\rangle_{,i}$, we finally find

\begin{equation}\label{eq: Fisher bispectrum}
\boldsymbol{F}_{ij} = \sum\limits_\nu \sum\limits_{\ell_1\ell_2\ell_3} \frac{\mu^{\nu, \ell_1\ell_2\ell_3}_{,i}\mu^{\nu, \ell_1\ell_2\ell_3}_{,j}}{\Delta_{\ell_1\ell_2\ell_3}C^\nu_{\ell_1}C^\nu_{\ell_2}C^\nu_{\ell_3}}, 
\end{equation}
with
\begin{equation}
\mu^{\nu, \ell_1\ell_2\ell_3}_{,i} = 
\frac{\partial B^\text{NLG}_{\ell_1\ell_2\ell_3}(\nu)}{\partial\theta_i}+\frac{\partial B^\text{LISW}_{\ell_1\ell_2\ell_3}(\nu)}{\partial\theta_i}.
\end{equation}

We use equations \eqref{eq: Fisher power spectrum} and \eqref{eq: Fisher bispectrum} to compute the Fisher matrix for power spectrum and bispectrum observations respectively. 
The information from  both modes of analysis can be combined simply by adding the Fisher matrices as we assume  both statistical measures to be uncorrelated \citep{Takada2004}.

\subsection{LISW detection signal to noise}
Although we have seen in \ref{sec: Bispectrum Representation} that the LISW bispectrum signal can be significantly lower than that of the NLG bispectrum, considering shape differences between the contribution may allow a significant signal increase. 
In order to assess whether the LISW bispectrum signal is detectable by future experiments, we assume that, for $\alpha \equiv (\ell_1, \ell_2, \ell_3, m_1, m_2, m_3)$ obeying the triangle conditions, we observe a bispectrum $B_{\alpha}^\text{obs}$.
Suppose the shape of $B_{\alpha}^\text{LISW}$ is fixed and it can be distinguished from other contributions to the bispectrum. 
Then, we create and minimize 
\begin{equation}
\chi^2 = \sum\limits_{\alpha} \frac{\left(B_{\alpha}^\text{obs} - B_\alpha^{\text{th}}\right)^2}{\sigma_\alpha^2},
\end{equation} 
where
\begin{equation}
B_\alpha^{\text{th}} \equiv \mathcal{A} B_\alpha^{\text{LISW}}+\mathcal{B} B_\alpha^{\text{NLG}}, 
\end{equation}
for some amplitudes $\mathcal{A}$ and $\mathcal{B}$.
For simplicity, it is assumed that amplitude for non-linear gravity is known exactly, so it can be subtracted from the observations, such that
\begin{equation}
\chi^2 = \sum\limits_{\alpha} \frac{\left(\tilde{B}_{\alpha}^\text{obs} -\mathcal{A} B_\alpha^{\text{LISW}}\right)^2}{\sigma_\alpha^2},
\end{equation}
where $\tilde{B}^\text{obs}_\alpha \equiv B^\text{obs}_\alpha - \mathcal{B}B^\text{NLG}_\alpha$.
Minimising this function with respect to the LISW amplitude $\mathcal{A}$, we obtain an estimator
\begin{equation}
\hat{\mathcal{A}} = \frac{\sum\limits_\alpha \tilde{B}_{\alpha}^\text{obs}B_{\alpha}^\text{LISW}/\sigma^2_\alpha}{\sum\limits_\alpha \left(B_{\alpha}^\text{LISW}\right)^2/\sigma^2_\alpha},
\end{equation}
with a variance on the estimator given by
\begin{equation}
\sigma_{\hat{\mathcal{A}}}^2 = \frac{1}{\sum\limits_\alpha \left(B_{\alpha}^\text{LISW}\right)^2/\sigma^2_\alpha}.
\end{equation}
The variance on the bispectrum is computed in \cite{Spergel1999} as
\begin{equation}\label{eq: noise B}
\sigma^2_{\alpha} = \left\langle B_\alpha^2 \right\rangle - \left\langle B_\alpha \right\rangle^2 \simeq \Delta_\alpha C_{\ell_1}^{\nu,\text{tot}}C_{\ell_2}^{\nu,\text{tot}}C_{\ell_3}^{\nu,\text{tot}},
\end{equation}
where $\Delta_\alpha$ is 6, 2, or 1 when all $\ell$'s, two $\ell$'s or no $\ell$'s are the same respectively. The $C_\ell$ here denote the angular 21cm power spectrum including detector noise.

\begin{figure}
\includegraphics[width=0.45\textwidth]{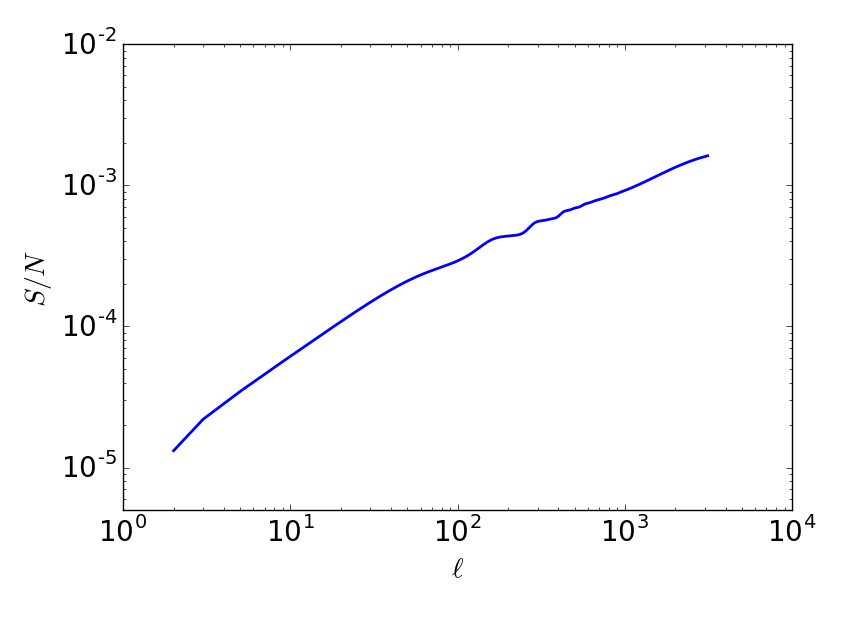}
\caption{Signal to noise for a LISW bispectrum signal detection vs $\ell$, the highest multipole moment observed, using MeerKAT in interferometry mode.}\label{Figure: SN}
\end{figure}

Assuming now that a fiducial value for our estimator is $\mathcal{A} = 1$, and that our estimator is unbiased, $\langle\hat{\mathcal{A}}\rangle = \mathcal{A}$,  we compute the signal to noise ratio for an IM experiment probing the LISW bispectrum,
\begin{equation}
\frac{S}{N} = \sqrt{\sum\limits_{\text{all }\alpha} \frac{(B_{\alpha}^\text{LISW})^2}{\sigma^2_\alpha}}.
\end{equation}
Importantly, we can sum out all $m$ indices by applying \eqref{eq: sum W3J squared over m}, such that 
\begin{equation}
\frac{S}{N} = \sqrt{\sum\limits_{\ell_1 \ell_2 \ell_3} \frac{(B_{\ell_1 \ell_2 \ell_3}^\text{LISW})^2}{\sigma^2_{\ell_1 \ell_2 \ell_3}}}.
\end{equation}
Fig. \ref{Figure: SN} shows the detection signal to noise ratio as a function of the largest $\ell$ mode included in the sum in the optimal case for which the shape of other contributing bispectra is known exactly.
Despite the large number of modes added, a direct detection of the LISW contribution to the bispectrum is impossible as even including information from small-scales does not increase the signal to noise ratio significantly above $10^{-3}$.
In comparison to the CMB, where the LISW contribution represents a major contaminant for primordial non-Gaussianity observations \citep{Kim2013,Planck2016NG}, the principal reason for the small S/N ratio found here, is that the power spectrum contribution to the noise term (see equation \eqref{eq: noise B}) is significantly larger due to the late stage of the gravitational growth of structure.  

Although a direct detection of the LISW bispectrum is impossible, ignoring it potentially biases the measurement of cosmological parameters from bispectrum observations \citep{Planck2016NG}.
We compute this bias term here \citep{Kim2004,Taylor2007}, but find values $\sim 8 - 10$ orders of magnitude lower than the expected errors, and thus the LISW effect insignificantly affects the 21cm bispectrum.

\section{Fisher Predictions} \label{sec: discussion}

\begin{table*}
\begin{center}
\caption{Marginal errors recovered from the Fisher forecasts. We show the results for $10^4$ hours of integration time for  bispectrum-only observations, power spectrum-only observations and the combined analysis. Each of these analyses is performed in interferometry and single-dish mode, and we finally combine the analysis to obtain the information gained over all scales.}
 \begin{tabular}{p{1.5cm} p{1.5cm} p{1.5cm} p{1.5cm} p{1.5cm} p{1.5cm} p{1.5cm} p{1.5cm} p{1.5cm} c} 
 \hline
 & & \multicolumn{3}{c}{\textbf{Interferometer}} & \multicolumn{2}{c}{\textbf{Single-Dish}} &  \multicolumn{2}{c}{\textbf{Combined}}\\
 \hline
 & & \centering CHIME & \centering MeerKAT & \centering SKA & \centering MeerKAT & \centering SKA & \centering MeerKAT & \centering SKA &\\
 \hline \hline
 \centering Parameter & \centering Fid. Value & \multicolumn{7}{c}{\textit{Marginalized error for bispectrum analysis}} \\  
 \hline\hline
 \centering $\Omega_\text{CDM}h^2$ & \centering 0.127 &  \centering $3.1\times 10^{-4}$  &  \centering $3.7\times 10^{-4}$  &  \centering $2.3\times 10^{-4}$  &  \centering $1.1\times 10^{-3}$  &  \centering $1.0\times 10^{-3}$  &  \centering $2.4\times 10^{-4}$  & \centering $1.7\times 10^{-4}$ &\\ 
 \hline
 \centering $\Omega_{\rm b} h^2$ &  \centering 0.022  &  \centering $8.2\times 10^{-5}$  &  \centering $1.0\times 10^{-4}$  &  \centering $6.0\times 10^{-5}$  &  \centering $3.0\times 10^{-4}$  &  \centering $2.9\times 10^{-4}$  &  \centering $8.4\times 10^{-5}$  & \centering $5.4\times 10^{-5}$ &\\
 \hline
\centering $\Omega_\Lambda$ &  \centering 0.684 &  \centering $2.1\times 10^{-4}$ &  \centering $2.4\times 10^{-4}$  &  \centering $1.7\times 10^{-4}$  &  \centering $1.7\times 10^{-3}$  &  \centering $1.6\times 10^{-3}$  &  \centering $2.3\times 10^{-4}$  & \centering $1.7\times 10^{-4}$ &\\
 \hline
 \centering $n_{\rm s}$ &  \centering 0.962 &  \centering  $5.7\times 10^{-4}$  &  \centering $5.7\times 10^{-4}$  &  \centering $3.9\times 10^{-4}$  &  \centering $2.5\times 10^{-3}$  &  \centering $2.3\times 10^{-3}$  &  \centering $3.8\times 10^{-4}$  & \centering $3.0\times 10^{-4}$ &\\ 
 \hline
 \centering $A_{\rm s}\times 10^9$ & \centering 1.562 &  \centering $5.9\times 10^{-3}$  &  \centering $7.2\times 10^{-3}$  &  \centering $4.5\times 10^{-3}$  &  \centering $2.1\times 10^{-2}$  &  \centering $1.9\times 10^{-2}$  &  \centering $4.2\times 10^{-3}$  & \centering $3.2\times 10^{-3}$ &\\
 \hline
 \centering $H_0$ &  \centering 67 &  \centering $8.0\times 10^{-2}$  &  \centering $9.7\times 10^{-2}$  &  \centering $5.6\times 10^{-2}$  &  \centering $3.0\times 10^{-1}$ &  \centering $2.8\times 10^{-1}$  &  \centering $7.0\times 10^{-2}$  & \centering $4.6\times 10^{-2}$ &\\
 \hline \hline
\centering Parameter & \centering Fid. Value &  \multicolumn{7}{c}{\textit{Marginalized error for power spectrum analysis}}\\
\hline \hline
 \centering $\Omega_\text{CDM}h^2$ & \centering 0.127 &  \centering  $9.3\times 10^{-4}$  &  \centering $4.9\times 10^{-4}$  &  \centering $3.8\times 10^{-4}$  &  \centering $2.5\times 10^{-3}$  &  \centering $1.4\times 10^{-3}$  &  \centering $4.6\times 10^{-4}$  & \centering $3.4\times 10^{-4}$ &\\ 
 \hline
 \centering $\Omega_{\rm b} h^2$ &  \centering 0.022  &  \centering  $4.2\times 10^{-4}$  &  \centering $2.9\times 10^{-4}$  &  \centering $2.4\times 10^{-4}$  &  \centering $1.0\times 10^{-3}$  &  \centering $6.5\times 10^{-4}$  &  \centering $2.5\times 10^{-4}$  & \centering $2.0\times 10^{-4}$ &\\
 \hline
\centering $\Omega_\Lambda$ &  \centering 0.684 &  \centering  $2.9\times 10^{-3}$  &  \centering $1.6\times 10^{-3}$  &  \centering $1.2\times 10^{-3}$  &  \centering $1.2\times 10^{-2}$  &  \centering $6.4\times 10^{-3}$  &  \centering $1.5\times 10^{-3}$  & \centering $1.1\times 10^{-3}$ &\\
 \hline
 \centering $n_{\rm s}$ &  \centering 0.962 &  \centering  $1.5\times 10^{-3}$  &  \centering $9.0\times 10^{-4}$  &  \centering $7.2\times 10^{-4}$  &  \centering $8.0\times 10^{-3}$  &  \centering $4.5\times 10^{-3}$  &  \centering $8.4\times 10^{-4}$  & \centering $6.5\times 10^{-4}$ &\\ 
 \hline
 \centering $A_{\rm s}\times 10^9$ & \centering 1.562 &  \centering  $8.4\times 10^{-3}$  &  \centering $8.4\times 10^{-3}$  &  \centering $6.3\times 10^{-3}$  &  \centering $4.9\times 10^{-2}$  &  \centering $2.5\times 10^{-2}$  &  \centering $7.6\times 10^{-3}$  & \centering $5.4\times 10^{-3}$ &\\
 \hline
 \centering $H_0$ &  \centering 67 &  \centering $3.0\times 10^{-1}$  &  \centering $2.3\times 10^{-1}$  &  \centering $1.9\times 10^{-1}$  &  \centering $8.9\times 10^{-1}$  &  \centering $5.1\times 10^{-1}$  &  \centering $2.1\times 10^{-1}$  & \centering $1.6\times 10^{-1}$ &\\
 \hline \hline
\centering Parameter & \centering Fid. Value & \multicolumn{7}{c}{\textit{Marginalized error for combined power spectrum + bispectrum analysis}} \\
\hline \hline
  \centering $\Omega_\text{CDM}h^2$ & \centering 0.127 &  \centering $9.2\times 10^{-5}$  &  \centering $1.2\times 10^{-4}$  &  \centering $5.5\times 10^{-5}$  &  \centering $6.5\times 10^{-4}$   &  \centering $4.9\times 10^{-4}$  &  \centering $1.1\times 10^{-4}$  & \centering $5.3\times 10^{-5}$ &\\ 
 \hline
 \centering $\Omega_{\rm b} h^2$ &  \centering 0.022  &  \centering $4.0\times 10^{-5}$  &  \centering $5.8\times 10^{-5}$  &  \centering $3.1\times 10^{-5}$  &  \centering $2.1\times 10^{-4}$   &  \centering $1.7\times 10^{-4}$  &  \centering $5.3\times 10^{-5}$  & \centering $2.9\times 10^{-5}$ &\\
 \hline
\centering $\Omega_\Lambda$ &  \centering 0.684 &  \centering $1.7\times 10^{-4}$  &  \centering $2.2\times 10^{-4}$  &  \centering $1.4\times 10^{-4}$  &  \centering $1.7\times 10^{-3}$   &  \centering $1.4\times 10^{-3}$  &  \centering $2.2\times 10^{-4}$  & \centering $1.4\times 10^{-4}$ &\\
 \hline
 \centering $n_{\rm s}$ &  \centering 0.962 &  \centering $1.4\times 10^{-4}$  &  \centering $1.1\times 10^{-4}$  &  \centering $8.2\times 10^{-5}$  &  \centering $1.2\times 10^{-3}$   &  \centering $8.1\times 10^{-4}$  &  \centering $1.0\times 10^{-4}$  & \centering $7.3\times 10^{-5}$ &\\ 
 \hline
 \centering $A_{\rm s}\times 10^9$ & \centering 1.562 &  \centering $1.7\times 10^{-3}$  &  \centering $2.2\times 10^{-3}$  &  \centering $1.2\times 10^{-3}$ &  \centering $1.3\times 10^{-2}$   &  \centering $9.8\times 10^{-3}$  &  \centering $1.9\times 10^{-3}$  & \centering $1.1\times 10^{-3}$ &\\
 \hline
 \centering $H_0$ &  \centering 67 &  \centering $3.2\times 10^{-2}$  &  \centering $4.5\times 10^{-2}$  &  \centering $2.2\times 10^{-2}$  &  \centering $1.9\times 10^{-1}$  &  \centering $1.6\times 10^{-1}$  &  \centering $4.1\times 10^{-2}$ & \centering $2.1\times 10^{-2}$ &\\
 \hline \hline
\end{tabular}
\label{table: statistics summary}
\end{center}
\end{table*}

\begin{figure*}
	\includegraphics[width=0.9\textwidth]{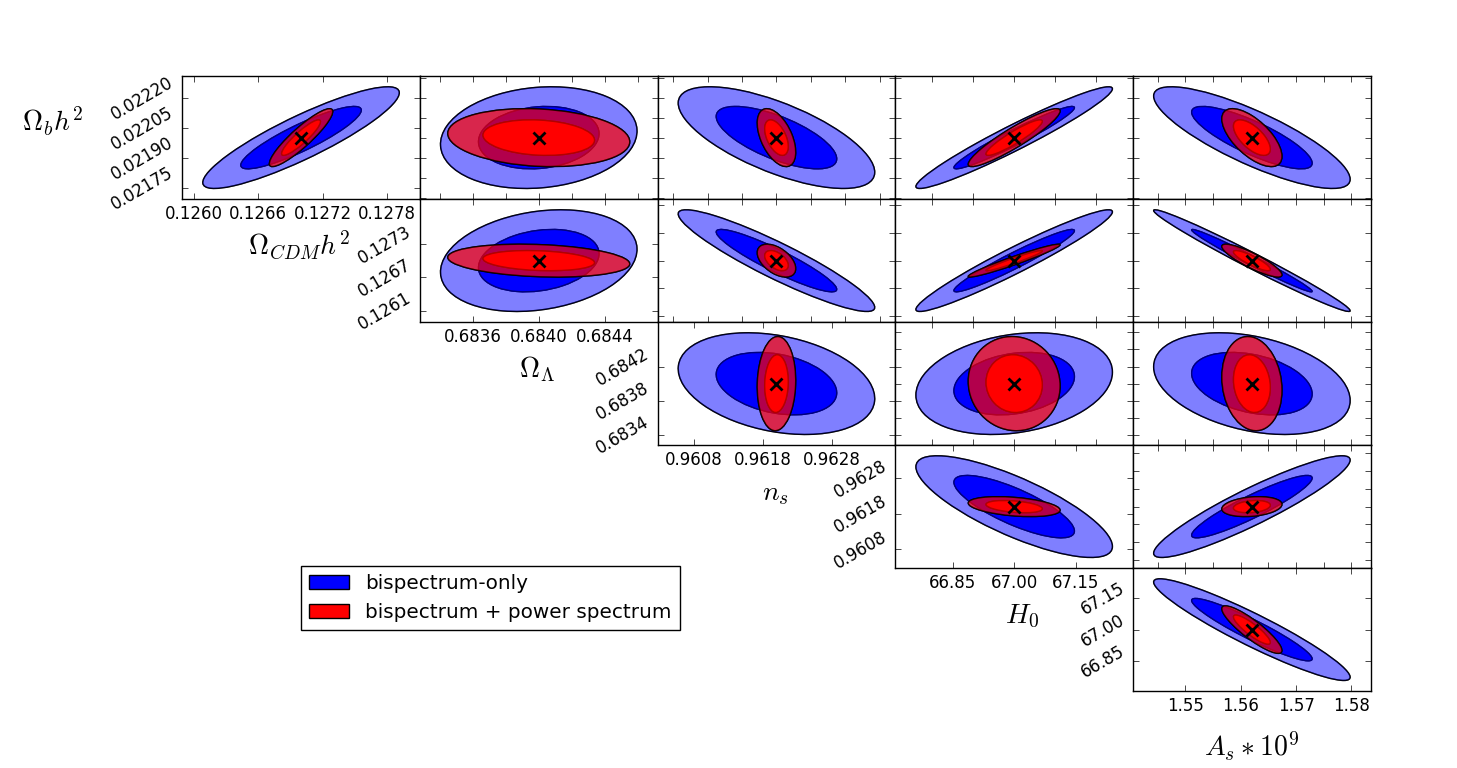}  
  \caption{Fisher forecasts for bispectrum-only (blue ellipses) and power spectrum + bispectrum (red ellipses) observations for MeerKAT in interferometry mode. We show the 68\% and 95\% credibility intervals for the cosmological parameters used in our analysis for a total integration time of $10^4$ hours. The cross shows the fiducial value of the parameters, and uniform priors are assumed.}\label{fig: ellipses}
\end{figure*}

Computing equations \eqref{eq: Fisher power spectrum} and especially \eqref{eq: Fisher bispectrum} is expensive due to the large number of modes, frequency bins and parameter combinations necessary to compute.
For the bispectrum, the number of modes scales as $\sim \ell_\text{max}^3$, meaning that bispectra containing small angular scales require an extraordinarily large number of modes to be computed.
The limiting scale at a given frequency is set by maximal baseline in the interferometer or the dish size of the single-dish observation.
In order for the bispectrum computation to be practical, we ignore any modes dominated by noise ($\ell_\text{max}\gtrsim 1600$), and thus impose an upper bound of $\ell_\text{max}=\min\limits(2\pi D_\text{tel/dish}/\lambda, 1600)$.
The method is therefore insensitive to contributions from scales smaller than $\sim 3\; {\rm Mpc}$, where complex gas physics gives rise to highly non-Gaussian objects which would affect the bispectrum at large $\ell$.
Additionally, the minimal baseline for an interferometer, $D_\text{min}$, sets the largest observable mode and thus we impose an $\ell_\text{min} = 2\pi D_\text{min} / \lambda$ for observations performed in interferometry mode.
As a consequence, interferometric studies of the 21cm bispectrum will not be sensitive to squeezed triangle configurations, and thus will not contain information from triangles which maximize its amplitude.
Further, the Fisher information varies smoothly as a function of $\ell_\text{max}$ and we therefore compute the statistical quantities with a step-size of $\Delta\ell_\text{max} = 80$ and linearly interpolate when summing over $\ell$ in equation \eqref{eq: Fisher bispectrum}. 

The results for the Fisher analysis are shown in table \ref{table: statistics summary}.
The table is subdivided into three sections, comparing the bispectrum-only observations for each of our fiducial experiments in the first seven rows. 
The next seven rows show the results for our power spectrum-only observations, before we combine both results in the final section of the table.
Furthermore, we compare results from observations made in interferometry and single-dish mode, and combine their Fisher matrices in the final column as they are uncorrelated.

We find that bispectrum observations have the potential to improve the parameter constraints from power spectrum observations significantly due to the large number of accessible modes, $N_\text{modes} \sim \ell_\text{max}^3$, but the bispectrum does not contain all the information and a combination of both statistics is required to obtain the best constraints.
All experiments exhibit errors from the bispectrum forecasts which are a factor of $\sim 1.1 - 7$ better than compared to the power spectrum.
Combinations of both show an order of magnitude reduction in our error forecasts for most parameters.
We find the strongest constraints across our analysis for $\Omega_\Lambda$, $n_{\rm s}$ and $H_0$.
This is in line with the expectation that IM experiments should improve the constraints of $H_0$ and $\Omega_\Lambda$ the most, as well as parameters such as $n_{\rm s}$ which are correlated to these \citep{Bull2015}.
Comparing our results for MeerKAT in interferometry mode to \cite{Planck2016a}, the 21cm power spectrum forecasts show a factor of $\sim 2-5$ decrease in marginalized errors for most parameters, with the exception of $\Omega_{\rm b} h^2$ where we do not achieve the same level of sensitivity. 
The bispectrum promises to tighten constraints on all cosmological parameters by up to a factor of 10, thus having the potential to bridge the gap between current low-redshift and CMB observations of the cosmological parameters.
Even in single-dish mode, we find that our power spectrum forecasts result in similar errors as those observed with \textit{Planck}, and the bispectrum again improving these findings typically by a factor of 3.
The best possible constraints are achieved by combining interferometric and single-dish observations of both the power spectrum and the bispectrum, these combinations marginally improve the constraints obtained from interferometric power spectrum and bispectrum combinations.

Of note is that we find that CHIME achieves better constraints from bispectrum observations than MeerKAT, even though power spectrum observations find error constraints of a factor of $\sim 2$ worse than MeerKAT. 
We find the CHIME noise power spectrum to be an order of magnitude lower than that of MeerKAT and would thus naively expect the CHIME power spectrum observations to result in stronger constraints. 
This is not the case as the 21cm power spectrum peaks on scales $\ell \sim 800$ which are on the edge of resolvability for CHIME.
Thus despite higher instrumental noise, compared to CHIME, MeerKAT is able to resolve smaller scales, due to its larger baselines, and thus is sensitive to the peak in the signal power spectrum.
For bispectrum observations, neither telescope is sensitive to the largest amplitude triangles and despite being sensitive to a larger number of modes, they are noisier for MeerKAT observations, such that CHIME is able to use the bispectrum to a higher potential.

Fig. \ref{fig: ellipses} shows the $1\sigma$ and $2\sigma$ error ellipses from our analysis for both bispectrum-only and power spectrum plus bispectrum combined observations by MeerKAT in interferometry mode.
The combination of the information gain from both bispectrum and power spectrum is thus not only useful to decrease errors, but can be a helpful tool to break degeneracies between parameters.

\section{Conclusions}\label{sec: conclusion}

The most precise observations of the CMB to date \citep{Planck2016a} have confirmed the simple picture of a six parameter cosmological model with a cosmological constant and a flat curvature. 
Although there is not sufficient evidence to strongly favour any other model at this time \citep{Heavens2017}, tensions between CMB and low-redshift observations of weak lensing and local measurements of the Hubble rate still persist.
New low-redshift probes may help to rectify these short-comings of the model and give new insights into the cosmological evolution since the time of recombination.
The cosmological 21cm signal is an ideal probe as HI and thus the 21cm signal is present at all epochs after the CMB is released.
21cm intensity mapping experiments will soon supplement galaxy surveys for mapping the large scale structure of the universe by observing the diffuse 21cm emission from hydrogen gas inside low-redshift galaxies.
These experiments will probe unprecedented cosmological volumes and  provide precise redshift information for their observations, due to the direct relation between the observed frequency of the signal and the redshift of the source.

We have studied the 21cm bispectrum and power spectrum in the context of IM observations by CHIME, MeerKAT, and SKA-mid and derived the expression for the 21cm bispectrum due to the non-linear collapse of structure post reionization.
For the first time, we derived the expected contribution to the 21cm bispectrum from the lensing-ISW bispectrum which is due to the evolution of the density field along the line of sight.
In contrast to CMB observations, we find, as expected, that the lensing-ISW bispectrum only introduces a negligible bias to the parameter constraints and we predict a cumulative signal to noise ratio of $10^{-3}$, making a detection impossible.
We introduce a new way of visualizing the bispectrum which allows for a direct relation between the triangle shape and the resulting amplitude.
Finally, we analysed the predictive capabilities of these bispectrum contributions in the context of a Fisher forecast model and found that the bispectrum from IM experiments has the potential to greatly improve cosmological parameter contrains. 
Although not sensitive to the largest amplitude triangles, the large number of observable modes should allow interferometric IM experiments to extract enough information to decrease parameter errors by an order of magnitude compared to the \textit{Planck} measurements.
For the best case scenario, the combined analysis of interferometry and single-dish observations of both power spectrum and bispectrum with SKA-mid, an impressive level of precision can be achieved.
We find a relative marginalized error of $< 0.1\%$ for all cosmological parameters, except for $\Omega_{\rm b}h^2$ for which we find a relative error of $\sim 0.13\%$. 
The bispectrum is especially sensitive to $n_{\rm s}$ where we find a relative marginalized error of $< 0.01 \%$.
It is important to reiterate that these results are heavily subjected to the level of foreground contamination and thus the level at which they can be removed.
We have used a simple foreground model and assumed an optimistic foreground removal efficiency of $\epsilon = 10^{-6}$ to explore the full, noise limited potential of bispectrum observations. 

\section*{Acknowledgements}
We would like to thank Suman Majumdar, Catherine Watkinson, Phil Bull and Alkistis Pourtsidou for their helpful conversations and suggestions.
CJS acknowledges the National Research
Fund, Luxembourg grant `Analytic and numerical analysis of the cosmic 21cm signal'.
JRP is pleased to acknowledge support from the European Research Council under ERC grant number 638743-FIRSTDAWN.


\appendix

\section{Limber Approximation}\label{Appx: Limber}
In the Limber approximation, for large $\ell$, Bessel functions are taken to be sharply peaked and are approximated by a Dirac delta function \citep{Loverde2008}, such that
\begin{equation}\label{eq: limber}
\begin{aligned}
\int dk k^2 f(k)& j_\ell[k r(z)]j_\ell[k q(z')]\\
&\simeq f\left[\frac{\ell+1/2}{r(z)}\right] \frac{\pi}{2r^2(z)}\frac{\delta^{\rm D}(z-z')}{|r'(z)|}.
\end{aligned}
\end{equation} 
In order to integrate equation \eqref{eq: NLG integral}, we use the Limber approximation and look at each $\ell$ term in turn.

\subsection{The $\ell = 0$ case}
We begin with the $\ell = 0$ term in \eqref{eq: NLG integral}, which can be written as
\begin{equation}
B_{12}^{\ell = 0} = A_{\ell_1\ell_2\ell_3}  b_\text{HI} \int dz D^2_+(z)  \delta\bar{T}_{\rm b}(z) W_\nu(z) \theta_{\ell_1}(z) \theta_{\ell_2}(z),
\end{equation}
where we have defined
\begin{equation}\label{eq: A0}
\begin{aligned}
A_{\ell_1\ell_2\ell_3} = &\frac{16}{\pi}\sqrt{\frac{(2\ell_1 + 1)(2\ell_2 + 1)(2\ell_3 + 1)}{(4 \pi)^3}}(2\ell_1+1)\\ 
&(2\ell_2+1)B_0 
\begin{Bmatrix}
\ell_1 & \ell_2 & \ell_3 \\
\ell_2 & \ell_1 & 0
\end{Bmatrix} 
 \begin{pmatrix}
\ell_1 & \ell_1 & 0 \\
0 & 0 & 0
\end{pmatrix} \\
& \begin{pmatrix}
\ell_2 & \ell_2 & 0 \\
0 & 0 & 0
\end{pmatrix}
\begin{pmatrix}
\ell_3 & \ell_1 & \ell_2 \\
0 & 0 & 0
\end{pmatrix},
\end{aligned}
\end{equation}
and
\begin{equation}
\theta_\ell(z) = b_\text{HI} \int dz' D_+(z') \delta\bar{T}_{\rm b}(z') W_\nu(z')\beta_\ell(z,z'),
\end{equation}
with
\begin{equation}
\beta_\ell(z,z') = \int dk k^2 P(k) j_\ell[k r(z)]j_\ell[k r(z')].
\end{equation}
Applying the Limber approximation \eqref{eq: limber} to $\beta_\ell$ gives
\begin{equation}\label{eq: beta limber 0}
\beta_\ell(z,z') \simeq \frac{\pi}{2r^2(z) r'(z)}P\left[\frac{\ell+1/2}{r(z)}\right] \delta^{\rm D}(z-z'),
\end{equation}
such that
\begin{equation}\label{eq: theta limber 0}
\theta_\ell(z) \simeq \frac{\pi b_\text{HI}}{2r^2(z) r'(z)}P\left[\frac{\ell+1/2}{r(z)}\right] D_+(z) \delta\bar{T}_{\rm b}(z) W_\nu(z).
\end{equation}

\subsection{The $\ell=1$ case}

For the $\ell=1$ case, we have that $\beta_1(k_1,k_2) = 2A_1\left(k_1/k_2 + k_2/k_1\right)$.
Therefore, \eqref{eq: NLG integral} contains two terms with $k$ integrals of the form, $\int d k_1 d k_2 k_1^3 k_2 \cdots$ and $\int d k_1 d k_2 k_1 k_2^3\cdots$. 
Defining functions similar to the $\ell = 0$ case, we find 
\begin{equation}\label{eq: l=1 case}
\begin{aligned}
B_{12}^{\ell=1} =&b_\text{HI} \sum\limits_{\ell'\ell''} A_{\ell_1\ell_2\ell_3}^{\ell'\ell''}\int dz W_\nu(z)  T_{\rm b}(z)D^2_{+}(z)\\
& \left[\theta^1_{\ell_1\ell'}(z)\theta^{-1}_{\ell_2\ell''}(z)+\theta^{-1}_{\ell_1\ell'}(z)\theta^1_{\ell_2\ell''}(z)\right],
\end{aligned}
\end{equation}
where we define
\begin{equation}\label{eq: A}
\begin{aligned}
A_{\ell_1\ell_2\ell_3}^{\ell'\ell''} = &-\frac{16}{\pi}\sqrt{\frac{(2\ell_1 + 1)(2\ell_2 + 1)(2\ell_3 + 1)}{(4 \pi)^3}}(2\ell'+1)\\
&(2\ell''+1)i^{\ell_1+\ell_2+\ell'+\ell''} 2A_1 \begin{Bmatrix}
\ell_1 & \ell_2 & \ell_3 \\
\ell'' & \ell' & 1
\end{Bmatrix}\\
&\begin{pmatrix}
\ell_1 & \ell' & 1 \\
0 & 0 & 0
\end{pmatrix}
\begin{pmatrix}
\ell_2 & \ell'' & 1 \\
0 & 0 & 0
\end{pmatrix}
\begin{pmatrix}
\ell_3 & \ell' & \ell'' \\
0 & 0 & 0
\end{pmatrix},
\end{aligned}
\end{equation}
and
\begin{equation}
\theta^q_{\ell\ell'}(z) = b_\text{HI}\int dz' D_+(z') \delta\bar{T}_{\rm b}(z') W_\nu(z')\beta^q_{\ell\ell'}(z,z'),
\end{equation}
with
\begin{equation}\label{eq: beta q}
\beta^q_{\ell\ell'}(z,z') = \int dk k^{2+q} P(k) j_\ell[k r(z)]j_{\ell'}[k r(z')].
\end{equation}
Importantly, the Wigner symbols in \eqref{eq: A} reduce the sum in \eqref{eq: l=1 case} to 4 terms, which all incidentally render the powers of $i$ even.
Only terms with $\ell' = \ell_1-1, \ell_1+1$ and $\ell'' = \ell_2-1, \ell_2+1$, as shown in the table below are non-zero.

\begin{center}
\begin{tabular}{l | l | l} 
 $\ell'  /  \ell'' $ & $\ell_2 -1$ & $\ell_2 + 1$\\ 
 \hline\hline
 $\ell_1-1$ & $\ell_1-1$, $\ell_2 -1$ & $\ell_1 -1$, $\ell_2 +1$ \\ 
 \hline
 $\ell_1+1$ & $\ell_1 +1$, $\ell_2 -1$ & $\ell_1 +1$, $\ell_2 +1$\\
 \hline
\end{tabular}
\end{center}
The difference between the $\ell$ indices in \eqref{eq: l=1 case} is one, such that we approximate $\ell\pm1 \sim \ell$ for Bessel function indices here.
We find this approximation to work well as most of the signal comes from large $\ell$-modes.
We thus apply the Limber approximation \eqref{eq: limber} with $f(k) = k^q P(k)$ to \eqref{eq: beta q}, and find
\begin{equation}
\beta^q_{\ell\ell'}(z,z') \simeq \frac{\pi(\ell+1/2)^q}{2r^{2+q}(z) r'(z)}P\left[\frac{\ell+1/2}{r(z)}\right] \delta^{\rm D}(z-z'),
\end{equation}
such that
\begin{equation}\label{eq: theta l1}
\theta^q_{\ell\ell'}(z)\simeq \frac{\pi b_\text{HI}(\ell+1/2)^q }{2r^{2+q}(z) r'(z)}P\left[\frac{\ell+1/2}{r(z)}\right] D_+(z) \delta\bar{T}_{\rm b}(z) W_\nu(z).
\end{equation}

\subsection{The $\ell=2$ case}
Similar to the $\ell=0$ case, $B_2$ is independent of $k$, and thus we can write
\begin{equation}
\begin{aligned}\label{eq: l=2 case}
B_{12}^{\ell=2} =& b_\text{HI}\sum\limits_{\ell'\ell''} A_{\ell_1\ell_2\ell_3}^{\ell'\ell''}\int dz W_\nu(z)  T_{\rm b}(z)D^2_{+}(z)\\
& \theta_{\ell_1\ell'}(z)\theta_{\ell_2\ell''}(z),
\end{aligned}
\end{equation}
where we define
\begin{equation}\label{eq: A2}
\begin{aligned}
A_{\ell_1\ell_2\ell_3}^{\ell'\ell''} = &\frac{16}{\pi}\sqrt{\frac{(2\ell_1 + 1)(2\ell_2 + 1)(2\ell_3 + 1)}{(4 \pi)^3}}(2\ell'+1)\\
&(2\ell''+1)\beta_2 i^{\ell_1+\ell_2+\ell'+\ell''}\begin{Bmatrix}
\ell_1 & \ell_2 & \ell_3 \\
\ell'' & \ell' & 2
\end{Bmatrix}\\
&\begin{pmatrix}
\ell_1 & \ell' & 2 \\
0 & 0 & 0
\end{pmatrix}
\begin{pmatrix}
\ell_2 & \ell'' & 2 \\
0 & 0 & 0
\end{pmatrix}
\begin{pmatrix}
\ell_3 & \ell' & \ell'' \\
0 & 0 & 0
\end{pmatrix},
\end{aligned}
\end{equation}
and
\begin{equation}
\theta_{\ell\ell'}(z) = b_\text{HI}\int dz' D_+(z') \delta\bar{T}_{\rm b}(z') W(z')\beta_{\ell\ell'}(z,z'),
\end{equation}
with
\begin{equation}
\beta_{\ell\ell'}(z,z') = \int dk k^{2} P(k) j_\ell[k r(z)]j_{\ell'}[k r(z')].
\end{equation}
Similar to the $\ell=1$ case, the Wigner symbols in \eqref{eq: A2} reduce the sum in \eqref{eq: l=2 case} to 9 non-zero terms, which all result in even powers of $i$. 
The terms are non-zero for combinations of $\ell' = \ell_1-2,\ell_1,\ell_1+2$ and $\ell'' = \ell_2-2,\ell_2,\ell_2+2$ as shown below.

\begin{center}
\begin{tabular}{|| l | l | l | l ||} 
 $\ell'  /  \ell'' $ & $\ell_2 - 2$ & $\ell_2$ & $\ell_2 + 2$\\ 
 \hline\hline
 $\ell_1-2$ & $\ell_1-2$, $\ell_2 -2$ & $\ell_1-2$, $\ell_2$ & $\ell_1 -2$, $\ell_2 +2$\\ 
 \hline
 $\ell_1$ & $\ell_1$, $\ell_2 -2$ & $\ell_1$, $\ell_2$& $\ell_1$, $\ell_2 +2$\\
 \hline
 $\ell_1+2$ & $\ell_1+2$, $\ell_2-2$ & $\ell_1+2$, $\ell_2$ & $\ell_1 -1$, $\ell_2 +1$\\
 \hline
\end{tabular}
\end{center}
Although at large $\ell$, we have $\ell \pm 2 \sim \ell$, we find that this approximation does not give robust results when applying the Limber approximation. 
Instead we assume that $P(k)$ varies slowly across the range of the peaks of both Bessel functions such that we can effectively evaluate it at either peak.
Similarly, we assume that $D_+\delta\bar{T}_{\rm b}$ varies slowly across the window, such that we may evaluate it at the window centre.
Hence, for combinations involving $\ell_1 \pm 2$ and $\ell_2 \pm 2$, we have
\begin{equation}\label{eq: thetallpm2}
\begin{aligned}
\theta_{\ell\ell\pm 2}(z) \simeq &\frac{\pi b_\text{HI}}{2 r^2(z)|r'(z)|}P\left[\frac{\ell+1/2}{r(z)}\right]D_+(z)\delta\bar{T}_{\rm b}(z)  \\
&\int dz'dk k^2 j_\ell[kr(z)] j_{\ell\pm 2}[k r(z')]W_\nu(z')\\
&\times \frac{2 r^2(z')|r'(z')|}{\pi}.
\end{aligned}
\end{equation}
We need to include the factor of $2 r^2(z')|r'(z')|/\pi$ into the integral, as evaluating $P(k)$ at the peak of the Bessel function introduces the inverse term when setting the $k$-integral to a delta function, and since we are evaluating the integral exactly here, we need to cancel out this normalization.
When $\ell' = \ell_1$ and $\ell'' = \ell_2$, we apply \eqref{eq: limber} similarly to the $\ell=0$ case, and recover \eqref{eq: beta limber 0} and \eqref{eq: theta limber 0}.


\section{Lensing coefficient derivation}\label{Appx: lensing coefficient}
The brightness temperature fluctuations projected onto the sky are perturbed along the line of sight by the ISW effect, and in angle by gravitational lensing,
\begin{equation}
\begin{aligned}
\delta T^\text{obs}_{\rm b}(\hat{\boldsymbol{n}},\nu) = &\delta T^\text{obs}_{\rm b,0}(\hat{\boldsymbol{n}},\nu) + \nabla\delta T^\text{obs}_{\rm b,0}(\hat{\boldsymbol{n}},\nu)\cdot\nabla\theta(\hat{\boldsymbol{n}},\nu)\\ 
&+ \nu \frac{d\delta T^\text{obs}_{\rm b,0}}{d\nu}(\hat{\boldsymbol{n}},\nu) \frac{\Delta\nu}{\nu}(\hat{\boldsymbol{n}},\nu),
\end{aligned}
\end{equation}
where the 0-index indicates the unperturbed field. 
These fluctuations can then be transformed into harmonic space,
\begin{equation}\label{eq: full alm}
\begin{aligned}
a_{\ell m}^\nu = &\int d^2 \hat{\boldsymbol{n}}Y_{\ell m}(\hat{\boldsymbol{n}}) \left[\delta T^\text{obs}_{\rm b,0}(\hat{\boldsymbol{n}},\nu) + \nabla\delta T^\text{obs}_{\rm b,0}(\hat{\boldsymbol{n}},\nu)\cdot\nabla\theta(\hat{\boldsymbol{n}},\nu) \right. \\ 
&\left.+ \nu \frac{d\delta T^\text{obs}_{\rm b,0}}{d\nu}(\hat{\boldsymbol{n}},\nu) \frac{\Delta\nu}{\nu}(\hat{\boldsymbol{n}},\nu)\right].
\end{aligned}
\end{equation}
We can separate out each term in equation \eqref{eq: full alm}. 
Then, according to eq (32), we define
\begin{equation}
a_{\ell m}^{\rm L,\nu} = \int d^2\hat{\boldsymbol{n}} Y_{\ell m}(\hat{\boldsymbol{n}}) \nabla\delta T^\text{obs}_{\rm b,0}(\hat{\boldsymbol{n}},\nu)\cdot\nabla\theta(\hat{\boldsymbol{n}},\nu),
\end{equation}
with
\begin{equation}
\theta(\hat{\boldsymbol{n}},\nu) = \sum\limits_{\ell'm'} \theta_{\ell'm'}^\nu Y^*_{\ell' m'}(\hat{\boldsymbol{n}}),
\end{equation}
and
{\begin{equation}
\delta T^\text{obs}_{\rm b,0}(\hat{\boldsymbol{n}},\nu) = \sum\limits_{\ell'm'} a_{\ell'm'}^{0,\nu} Y^*_{\ell' m'}(\hat{\boldsymbol{n}}).
\end{equation}
We thus find
\begin{equation}
\begin{aligned}
a_{\ell m}^{\rm L,\nu} &= \sum\limits_{\ell'\ell''m'm''}\int d^2\hat{\boldsymbol{n}} a_{\ell'm'}^{*0,\nu}\theta^{*\nu}_{\ell''m''} \\
&\times Y^*_{\ell m}(\hat{\boldsymbol{n}})\nabla Y^*_{\ell' m'}(\hat{\boldsymbol{n}})\cdot\nabla Y^*_{\ell'' m''}(\hat{\boldsymbol{n}}),
\end{aligned}
\end{equation}
Where we have used the fact that the fluctuations are real.
Further, one can use the properties of the spherical harmonics and the following identity for functions $A$, $B$, and $C$,
\begin{equation}
\int d\hat{\boldsymbol{n}} C\nabla A \cdot \nabla B = \frac{1}{2}\int d\hat{\boldsymbol{n}}\left(AB\nabla^2C - AC \nabla^2B - BC \nabla^2 A\right),
\end{equation}
to show that the angular integral becomes
\begin{equation}
 \int d\hat{\boldsymbol{n}}Y^*_{\ell m}(\hat{\boldsymbol{n}})\nabla Y^*_{\ell' m'}(\hat{\boldsymbol{n}})\cdot\nabla Y^*_{\ell'' m''}(\hat{\boldsymbol{n}}) = W_{\ell\ell'\ell''}^{mm'm''}, 
\end{equation}
where
\begin{equation}
W_{\ell\ell'\ell''}^{mm'm''}\equiv \frac{1}{2}(-1)^{m+m'+m''} L_{\ell\ell'\ell''} \mathcal{H}_{\ell\ell'\ell''}^{mm'm''},
\end{equation}
with
\begin{equation}
L_{\ell\ell'\ell''}\equiv -\ell(\ell+1) + \ell'(\ell'+1)+\ell''(\ell''+1).
\end{equation}
Therefore, we find
\begin{equation}
a_{\ell m}^{\rm L,\nu} = \sum\limits_{\ell' \ell''m'm''} W_{\ell\ell'\ell''}^{mm'm''} a_{\ell'm'}^{*0,\nu}\theta^{*\nu}_{\ell''m''}.
\end{equation}
The harmonic transforms can be related to the 3D fields via
\begin{equation}
a_{\ell m}^\nu = \int d^2\nhat \delta T_{\rm b}^{\text{obs}} (\hat{\boldsymbol{n}},\nu) Y_{\ell m}(\nhat),
\end{equation} 
where the closure relation for spherical harmonics can be applied to obtain 
\begin{equation}
\begin{aligned}
a_{\ell m}^\nu =  &4\pi i^\ell\int dz W_\nu(z) \delta \bar{T}_{\rm b}(z) b_\text{HI}(z)D_+(z)\\
& \times\int \frac{d^3\boldsymbol{k}}{(2\pi)^3} \tilde{\delta}(\boldsymbol{k})j_\ell[kr(z)]Y_{\ell m}(\hat{\boldsymbol{k}}).
\end{aligned}
\end{equation}
Here
\begin{equation}
\begin{aligned}
\theta^{\nu}_{\ell m} &= \int d^2\nhat \theta(\hat{\boldsymbol{n}},\nu) Y_{\ell m}(\nhat) \\
& = \int d^2\nhat dz W_\nu(z)\theta\left[r(z)\hat{\boldsymbol{n}},z\right] Y_{\ell m}(\nhat),
\end{aligned}
\end{equation}
with $\theta\left[r(z)\hat{\boldsymbol{n}},z\right]$ given by equation \eqref{eq: lensing potential},
\begin{equation}
\begin{aligned}
\theta^{\nu}_{\ell m} &= -\frac{2}{c^2}\int d^2\nhat dz W_\nu(z) Y_{\ell m}(\nhat) \times\\
& \int_0^{r(z)}dr'\frac{S_k\left[r(z)-r'\right]}{S_k\left[r(z)\right]S_k(r')}\Phi(r'\nhat).
\end{aligned}
\end{equation}

\section{LISW power spectrum}\label{Appx: LISW power}
Let us first write down an expression for the ISW coefficients. 
From \eqref{eq: full alm}, 
\begin{equation}
a_{\ell m}^{\text{ISW},\nu} = \int d^2\nhat  Y^*_{\ell m}(\nhat) \underbrace{\nu \frac{d\delta T^\text{obs}_{\rm b,0}}{d\nu}(\nhat)}_{\eta^\nu(\nhat)} \frac{\Delta\nu}{\nu}(\nhat,\nu).
\end{equation}
We relate the projection on the sky to the 3D field, 
\begin{equation}
\begin{aligned}
a_{\ell m}^{\text{ISW},\nu} &= \int d^2\nhat  Y^*_{\ell m}(\nhat)  \eta^\nu(\nhat) \int dz W_\nu(z)\frac{\Delta\nu}{\nu}\left[r(z)\nhat,z\right] \\
&= \frac{2}{c^3}\int d^2\nhat  dz Y^*_{\ell m}(\nhat)  \eta^\nu(\nhat) W_\nu(z)\\
& \times \int^{r(z)}_0 dr'\pd{\Phi}{t}(r'\nhat,z),
\end{aligned}
\end{equation}
where we assume $\eta^\nu(\nhat) = \eta(z) = \nu(z) \frac{d\bar{T}_{\rm b}}{d\nu}(z)$ to lowest order.

We then define $Q_\ell(\nu,\nu)$ via equation \eqref{eq: Ql def}. Applying the Kronecker deltas, we find
\begin{equation}
\begin{aligned}
Q_\ell(\nu,\nu) &= \left\langle -\frac{2}{c^2}\int d\nhat dz W_\nu(z)Y_{\ell m}(\nhat)\right. \\
&\times\int_0^{r(z)}dr' \frac{S_k[r(z)-r']}{S_k[r(z)]S_k(r')} \\
&\times\frac{2}{c^3}\int d\nhat' dz' W_\nu(z')Y_{\ell m}(\nhat')\eta(z') \\
&\left.\times\int_0^{r(z')}dr''\pd{\Phi}{t}(r''\nhat',z')\right\rangle.
\end{aligned} 
\end{equation} 
We then write $\Phi$ in terms of its Fourier transform and expand the exponential according to equation \eqref{eq: exp}. 
The resulting expression can be summed over using the spherical harmonics closure relations and through the definition of the power spectrum for the gravitational potential,
\begin{equation}
\left\langle\pd{\Phi}{t}(\boldsymbol{k},z)\Phi(\boldsymbol{k}',z')\right\rangle=\frac{(2\pi)^3}{2}\pd{P_\Phi}{t}(k, z, z') \delta^D(\boldsymbol{k}+\boldsymbol{k}'), 
\end{equation}
we find
\begin{equation}
\begin{aligned}
Q_\ell(\nu,\nu) &= \frac{2(4\pi)^2}{c^5}\int dz dz' W_\nu(z) W_\nu(z') \eta(z')\\
&\times\int_0^{r(z)}dr'\int_0^{r(z')}dr''\frac{S_k[r(z)-r']}{S_k[r(z)]S_k(r')} \\
&\times\int\frac{k^2dk}{(2\pi)^3}\pd{P_\Phi}{t}(k,z,z')j_\ell(kr')j_\ell(kr'').
\end{aligned} 
\end{equation} 
We then apply the Limber approximation (see Appendix \ref{Appx: Limber}), integrate out the delta function introduced, and change integration variable to obtain
\begin{equation}
\begin{aligned}
Q_\ell(\nu,\nu) &= \frac{2}{c^4}\int dz W_\nu(z)\eta(z) \int dz' W_\nu(z') \\
&\times\int_0^{z'} dz''\frac{S_k[r(z')-r(z'')]}{S_k[r(z')]S_k[r(z'')]r(z'')^2} \\
&\times\left.\frac{\partial P_\Phi}{\partial z}(k,z'')\right|_{k=\ell/r(z'')},
\end{aligned} 
\end{equation}
where
\begin{equation}
\eta(z) = -(1+z)\frac{d\delta\bar{T}_{\rm b}}{dz}(z),
\end{equation}
and 
\begin{equation}
P_\Phi(k,z) = \left(\frac{3}{2}\Omega_{\rm M,0}\right)^2 \left(\frac{H_0}{k}\right)^4 P(k,z) (1+z)^2.
\end{equation}
Finally, we assume that both $\nu$ and the integral of the power spectrum vary slowly over the width of the window, which results in equation \eqref{eq: ql definition}.

\section{Orthogonality relations of Wigner-3J symbol}
\label{Appx: wigner}
The Wigner-3J symbols obey the following orthogonality relation \citep{Sobelman1979}:
\begin{equation}
 \sum\limits_{m_1,m_2} \begin{pmatrix}
\ell_1 & \ell_2 & \ell_3 \\
m_1 & m_2 & m_3
\end{pmatrix}\begin{pmatrix}
\ell_1 & \ell_2 & \ell'_3 \\
m_1 & m_2 & m'_3
\end{pmatrix} = \frac{\delta_{\ell_3 \ell'_3} \delta_{m_3 m'_3}}{(2\ell_3+1)},
\end{equation}
for $\ell_1$, $\ell_2$ and $\ell_3$ obeying the triangle conditions.
From this result we find a corollary by summing over the last $m$,
\begin{equation}\label{eq: sum W3J squared over m}
\sum\limits_{m_1,m_2,m_3} \begin{pmatrix}
\ell_1 & \ell_2 & \ell_3 \\
m_1 & m_2 & m_3
\end{pmatrix}^2 = 1,
\end{equation}
where again the $\ell$ modes need to satisfy the triangle conditions, otherwise the sum is zero.



\bibliography{PaperBib}




\bsp	
\label{lastpage}
\end{document}